\documentclass[reprint,amsmath,amssymb,aps]{revtex4-2}

\usepackage[utf8]{inputenc}
\usepackage{mathtools}
\usepackage{natbib}
\usepackage{graphics}
\usepackage{color}
\usepackage{mathrsfs}
\usepackage{dcolumn}
\usepackage{bm}
\usepackage{floatrow}   
\usepackage{graphicx,xcolor} 
\usepackage[framemethod=tikz]{mdframed}
\usepackage{amssymb,amsmath,amsfonts,amstext}
\usepackage{stmaryrd}
\usepackage{verbatim}
\usepackage{amsthm}
\usepackage{esint}
\usepackage{caption}
\usepackage{cancel}
\usepackage{subcaption}
\usepackage{bm}
\usepackage[T1]{fontenc}
\graphicspath{ {./Figures/} }
\usepackage{makecell}

\newcommand{\K}{\mathcal{K}}

\begin{document}


\title{Statistical-Physics-Informed Neural Networks (Stat-PINNs): \\
A Machine Learning Strategy for Coarse-graining Dissipative Dynamics}



\author{Shenglin Huang$^1$}
\author{Zequn He$^1$}
\author{Nicolas Dirr$^2$}
\author{Johannes Zimmer$^3$}
\author{Celia Reina$^1$}
\email{creina@seas.upenn.edu}
\affiliation{$^1$Department of Mechanical Engineering and Applied Mechanics, University of Pennsylvania, Philadelphia, PA 19104, USA}
\affiliation{$^2$School of Mathematics, Cardiff University, Cardiff CF24 4AG, UK}
\affiliation{$^3$School of Computation, Information and Technology, Technische Universit\"at M\"unchen, Boltzmannstr. 3, 85748
Garching, Germany}

\begin{abstract}

Machine learning, with its remarkable ability for retrieving information and identifying patterns from data, has emerged as a powerful tool for discovering governing equations. It has been increasingly informed by physics, and more recently by thermodynamics, to further uncover the thermodynamic structure underlying the evolution equations, i.e., the thermodynamic potentials driving the system and the operators governing the kinetics. However, despite its great success, the inverse problem of thermodynamic model discovery from macroscopic data is in many cases non-unique, meaning that multiple pairs of potentials and operators can give rise to the same macroscopic dynamics, which significantly hinders the physical interpretability of the learned models. In this work, we propose a machine learning framework, named as Statistical-Physics-Informed Neural Networks (Stat-PINNs), which further encodes knowledge from statistical mechanics and resolves this non-uniqueness issue for the first time. The framework is here developed for purely dissipative isothermal systems. Interestingly, it only uses data from short-time particle simulations to learn the thermodynamic structure, which can in turn be used to predict long-time macroscopic evolutions. We demonstrate the approach for particle systems with Arrhenius-type interactions, common to a wide range of phenomena, such as defect diffusion in solids, surface absorption and chemical reactions. Our results from Stat-PINNs can successfully recover the known analytic solution for the case with long-range interactions and discover the hitherto unknown potential and operator governing the short-range interaction cases. We compare our results with an analogous approach that solely excludes statistical mechanics, and observe that, in addition to recovering the unique thermodynamic structure, statistical mechanics relations can increase the robustness and predictive capability of the learning strategy.

\end{abstract}

\maketitle

\section{Introduction}
\label{Sec:Intro}

\begin{figure*}
\centering
\includegraphics[width=0.7\linewidth]{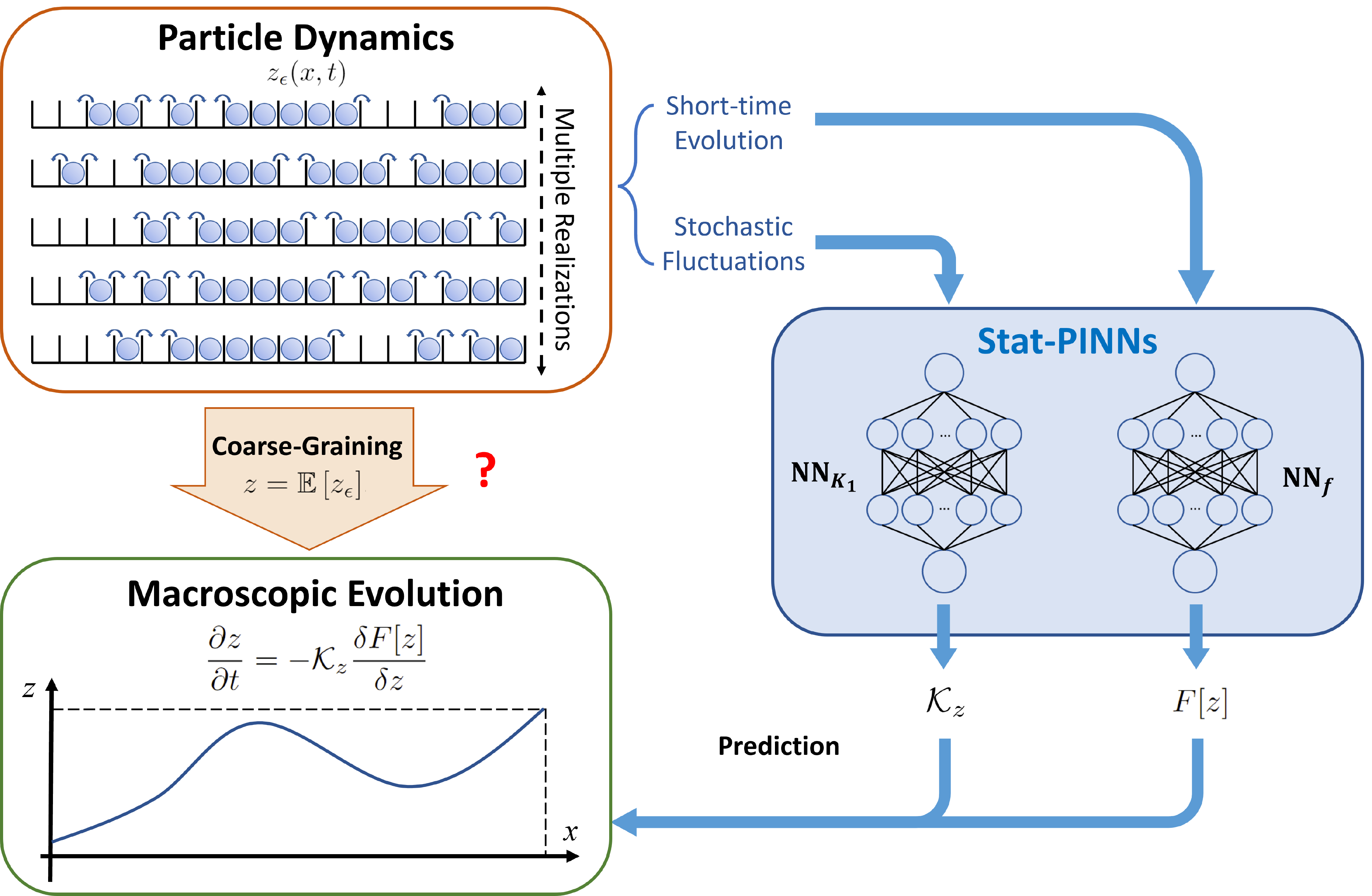}
\caption{Sketch of the Stat-PINNs (Statistical-Physics-Informed Neural Networks) framework for discovering the macroscopic evolution equation from short-time particle simulations.}
\label{fig:FlowChart}
\end{figure*}

Dissipative phenomena are pervasive across material systems, from diffusion in gases, to viscous flow in fluids, to plasticity in crystalline and granular media. Yet, our understanding of these phenomena is severely hindered by computational and theoretical challenges. Computationally, direct particle simulations remain elusive for macroscopic length- and time-scales, even with the latest supercomputers. And theoretically, only a small number  of particle systems enjoy an analytical macroscopic description \citep{kipnis1998scaling,Presutti2009a,bodineau2016brownian,katsoulakis2003coarse}. Notwithstanding, great progress has been made in the last 15 years in the understanding of the structure of the equations away from equilibrium. Indeed, the General Equation for Non-Equilibrium Reversible-Irreversible Coupling (GENERIC) formalism \citep{grmela1997dynamics,ottinger1997dynamics,ottinger2005beyond} provides the structure of macroscopic evolution equations as they emerge from underlying microscopic Hamiltonian dynamics while ensuring compatibility with the laws of thermodynamics. According to this formalism, the state variables $z$ describing a closed system  evolve through a combination of a symplectic operator acting on the energy of the system and a dissipative operator acting on the entropy.  
For isothermal systems, which are the focus of this article, the evolution reads
\begin{equation}
\label{Eq:T-GENERIC}
    \frac{\partial z}{\partial t} 
    =  -\K_z \frac{\delta F[z]}{\delta z},
\end{equation}
where $\K_z$ is a symmetric positive semi-definite operator that depends on the fields $z$ and $F$ is the free energy.
The GENERIC formalism has proven to be very advantageous from a modeling standpoint \citep{mielke2011formulation}, and its structure reveals additional information, such as the free energy as Lyapunov functional for systems evolving according to Eq.~\ref{Eq:T-GENERIC}. It is also endowed with strong statistical mechanics foundations \citep{ottinger2021framework,montefusco2021framework,li2019harnessing, kraaij2020fluctuation}, leading directly, for instance, to the stochastic PDE describing the evolution of finitely many particles (the so-called equation of fluctuating hydrodynamics).


 With the advent of machine learning, several strategies have emerged to discover the thermodynamic potentials and operators of the GENERIC equation. These include, in the context of ODEs, Structure-Preserving Neural Networks (SPNNs) \citep{hernandez2021deep}, GENERIC Neural Ordinary Differential Equations (GNODEs) \citep{lee2021machine} and GENERIC formalism Informed Neural Networks (GFINNs) \citep{zhang2022gfinns}. Within a variational perspective, Variational Onsager Neural Networks (VONNs) \citep{huang2022variational} learn the action density guiding the partial differential equations (of GENERIC type) in isothermal systems. Yet, all of the above methods can suffer from a lack of uniqueness, which commonly occurs, for example, in diffusive phenomena. This is already  the case for the simplest case of linear diffusion
\begin{equation*}
   \frac{\partial \rho}{\partial t} = \Delta \rho. 
\end{equation*}
This equation can be written in the form of Eq.~\ref{Eq:T-GENERIC} with $\K_\rho =1$ and $F = \frac 1 2 \int |\nabla \rho|^2 dx$ (the so-called $L^2$ gradient flow of the Dirichlet integral), or, alternatively, with $\K_{\rho} \xi= -\nabla \cdot (\rho \nabla \xi)$ and $F =  \int \rho \ln \rho\, \mbox{d} x$ (the so-called Wasserstein gradient flow of the Boltzmann entropy). 
This example shows that evolution equations can have non-unique GENERIC representations, despite the fact that thermodynamic potentials and operators are unique for a given particle system upon coarse-graining. In the example above, Brownian motion leads to the Wasserstein evolution and the Boltzmann entropy, while there is no natural particle foundation of the Dirichlet integral. This lack of uniqueness therefore severely limits the \emph{physics} learned by the machine learning algorithms if information from underlying particle models is not incorporated.

We here resolve this non-uniqueness issue for the first time by integrating statistical physics relations within the learning strategy.
While noise in the data is typically considered an undesirable feature, when such noise represents physical thermal fluctuations of particle data, it encodes a wealth of information that has proven instrumental in advancing our understanding of non-equilibrium phenomena \citep{kubo1966fluctuation,seifert2012stochastic,sevick2008fluctuation}. Of particular interest to this work is the profound link between fluctuations of a given variable and the true dissipative operator governing its evolution, via an infinite-dimensional fluctuation-dissipation relation \citep{ottinger2005beyond,li2019harnessing}. In particular, by measuring the fluctuations of $z$ in addition to its expected value, it is possible to uniquely determine the dissipative operator guiding the evolution. The proposed machine learning strategy, which we denote as Statistical-Physics Informed Neural Networks (Stat-PINNs), is schematically shown in Fig.~\ref{fig:FlowChart} for purely dissipative phenomena. It consists of two neural networks aimed at learning the unique thermodynamic dissipative operator and thermodynamic potential (the free energy $F$ for an isothermal process). Interestingly, both of these quantities, which govern the dynamics over arbitrarily long times, may be learned from short-time particle simulations.


Stat-PINNs is here developed in detail for conservative dissipative phenomena, such as diffusive processes, as their inverse problem is known to suffer from a lack of uniqueness. The proposed paradigm strongly encodes all the properties of the dissipative operator: acting linearly on the thermodynamic force (while generally being nonlinear in the state variables), being symmetric and positive semi-definite, as well as the restrictions associated to conservative fields $z$. We demonstrate the approach for particle systems with Arrhenius type interactions, which are very common in thermally activated processes such as defect diffusion in solids, surface absorption and chemical reactions. The specific particle process chosen has a known analytical solution for long-range interactions, while the continuum equation for short-range interactions is unknown, to the best of the authors' knowledge. Furthermore, it is one-dimensional, making it computationally feasible to perform long-time particle simulations as comparison data, to validate the proposed approach. Interestingly, Stat-PINNs is capable of recovering the analytical free energy and dissipative operator for the case of long-range interactions, as well as discover those for the case of short-range interactions. The results are compared to an analogous approach that solely includes thermodynamic relations and conservation laws, i.e., only excluding statistical mechanics. We observe that, in addition to recovering the unique thermodynamic potential and operator, statistical
mechanics relations increase the robustness and predictive capability of the learning strategy.

We remark that Stat-PINNs purposely uses a neural network to discover the free energy from data, in lieu of some alternative approaches rooted in statistical physics. One such alternative option would have been to use large deviation theory to determine the thermodynamic potential. While this approach is natural and elegant from a viewpoint of statistical mechanics, it requires resolution on an exponential scale, while the potential is here fitted directly. There is also a rich body of literature on the computation of equilibrium free energies (see~\cite{Lelievre2010a} for an in-depth presentation). These approaches allow the computation of free energy differences as required here, however only in a setting where a perturbative analysis of the system of interest is possible. The example of the Arrhenius process below demonstrates that the method presented here does not rely on perturbations.  

The paper is organized as follows. In Sec.~\ref{Sec:Theory}, we review the GENERIC formalism for isothermal dissipative systems and discuss the infinite-dimensional fluctuation dissipation relation needed to uniquely discover the dissipative operator. Next, in Sec.~\ref{Sec:Stat-PINNs}, we introduce the Stat-PINNs architecture and the corresponding structure-preserving parameterization. This machine learning strategy is then applied in Sec.~\ref{Sec:Arrhenius} over three Arrhenius-type interacting particle processes, of which only one of them has a known analytic solution. There, the results are compared to long-time particle simulations and physics-informed neural networks (PINNs) not informed by statistical mechanics. Finally, conclusions will be drawn in Sec.~\ref{Sec:Conclusion}.

\section{Isothermal dissipative dynamics}
\label{Sec:Theory}


We consider an isothermal system whose macroscopic evolution may be written as a gradient flow of the non-equilibrium free energy as in Eq.~\ref{Eq:T-GENERIC}.
There, $z=z(x,t)$ describes the field(s) of interest, and $F[z] = \beta \hat{F}[z]$ is the normalized total free energy functional (this will be referred to as free energy in the following, in the interest of simplicity), where $\hat{F}[z]$ is the original free energy functional in energy units, and $\beta = 1/k_B T$, with $k_B$ being the Boltzmann constant and $T$ the temperature of system. Further, $\delta F / \delta z$ is the functional derivative of the free energy, which acts as the thermodynamic driving force of the dissipative dynamics. Finally, $\mathcal{K}_z$ is a linear, symmetric and positive semi-definite operator, where the subscript $z$ emphasizes its dependency on the field(s) $z$, which can be nonlinear in general. Dissipative dynamics of the form of Eq.~\ref{Eq:T-GENERIC} were proposed by Ginzburg-Landau and may be seen as a special case of the GENERIC formalism for a system at constant temperature \citep{ottinger2005beyond,mielke2011formulation}.

The continuum equation \ref{Eq:T-GENERIC} can be mathematically seen as the system's dynamics in the limit of infinite number of particles. For finite, yet large number of particles, isothermal dissipative systems can often be described by a stochastic differential equation of the form (to be interpreted in  a suitable way as discussed below)
\begin{equation} 
\label{Eq:Sto_T-GENERIC-Klim}
    \frac{\partial z_\epsilon}{\partial t}
    = -\K_{z_\epsilon} \frac{\delta F[z_\epsilon]}{\delta z_\epsilon} 
     + \sqrt{2 \epsilon \mathcal{K}_{z_\epsilon}} \diamond \dot{W}_{x,t},
\end{equation}
where $\epsilon$ represents the level of ``zooming out'' in the description, and hence is proportional to the inverse of the average particle density.  Here $\dot{W}_{x,t}$ is a space-time white noise satisfying $\mathbb{E} [ \dot{W}_{x,t} ] = 0$ and $\mathbb{E} [ \dot{W}_{x,t} \dot{W}_{x',t'} ] = \delta(x-x')\delta(t-t')$, with $\mathbb{E}\left[\cdot\right]$ denoting the expectation and $\delta(\cdot)$ being the Dirac delta function. The stochastic integral denoted with the symbol $\diamond$ is the Klimontovich integral, see~\cite{ottinger2021framework}. From a physical viewpoint, this equation is natural, as the Gibbs measure $ \exp(-\frac 1 \epsilon F)$ is, at least formally, invariant under Eq.~\ref{Eq:Sto_T-GENERIC-Klim}, as we show below in a simple setting. The reformulation of Eq.~\ref{Eq:Sto_T-GENERIC-Klim} as an equation with It\^o noise results in 
\begin{equation} 
\label {Eq:Sto_T-GENERIC-Itoeps} 
    \frac{\partial z_\epsilon}{\partial t}
    = -\K_{z_\epsilon} \frac{\delta F[z_\epsilon]}{\delta z_\epsilon} 
     +\epsilon \partial_{z_\epsilon}\mathcal{K}_{z_\epsilon}   + \sqrt{2 \epsilon \mathcal{K}_{z_\epsilon}} \dot{W}_{x,t}.
\end{equation}
 We refer the reader to~\citep[Section~1.2.5]{ottinger2005beyond} for Eq.~\ref{Eq:Sto_T-GENERIC-Itoeps}.
The second but last term is of lower order, $O(\epsilon)$, and can thus be neglected in the computations, where it is sufficient to resolve $O(\sqrt{\epsilon})$. We thus study 
\begin{equation} 
\label{Eq:Sto_T-GENERIC}
    \frac{\partial z_\epsilon}{\partial t}
    = -\K_{z_\epsilon} \frac{\delta F[z_\epsilon]}{\delta z_\epsilon} 
       + \sqrt{2 \epsilon \mathcal{K}_{z_\epsilon}} \dot{W}_{x,t},
\end{equation}
where the noise is of It\^o form. We highlight that the fluctuation operator $\sigma_{z_\epsilon}=\sqrt{2 \epsilon \mathcal{K}_{z_\epsilon}}$ acting on the white noise $\dot{W}_{x,t}$ is related to the dissipative operator $\K_{z_\epsilon}$ through an infinite-dimensional fluctuation-dissipation relation, $\sigma_{z_\epsilon} \sigma_{z_\epsilon}^*=2\epsilon\mathcal{K}_{z_\epsilon}$. Existence of a solution to Eq.~\ref{Eq:Sto_T-GENERIC} is typically very subtle; for atomistic initial data as data representing particles as in this article, martingale solutions can be shown to exist for simple cases \citep{Renesse}. For regularization by correlated noise, as it is relevant for applications, existence results are available, see \cite{Fehrman2021a}. 

We sketch the invariance of the Gibbs measure $\frac 1 Z \exp(-\frac 1 \epsilon F)$ mentioned above, for Eq.~\ref{Eq:Sto_T-GENERIC-Itoeps} interpreted as an ordinary differential equation, 
\begin{equation} 
    \frac{\partial z_\epsilon}{\partial t}
    = -\K_{z_\epsilon}\nabla F
     +\epsilon \nabla\mathcal{K}_{z_\epsilon}  + \sqrt{2 \epsilon \mathcal{K}_{z_\epsilon}} \dot{W},
\end{equation}
where $W$ is a Wiener noise. The associated Fokker-Planck equation reads $\frac{\partial \rho}{ \partial t} = L^* \rho$, where the formal adjoint of the generator, $L^*$, is given by
\begin{equation}
\begin{split}
    L^* \rho 
    & = \nabla \cdot ( \mathcal{K}_{z_\epsilon} \rho \nabla F) - \epsilon \nabla \cdot (\nabla \mathcal{K}_{z_\epsilon} \rho) + \epsilon \nabla \cdot \nabla (\mathcal{K}_{z_\epsilon} \rho) 
    \\
    & = \nabla \cdot \left[( \mathcal{K}_{z_\epsilon} \rho \nabla F)  + \epsilon \mathcal{K}_{z_\epsilon} \nabla \rho\right],
\end{split}
\end{equation}
and this expression vanishes for $\rho = \frac 1 Z \exp (- \frac 1 \epsilon F)$.

\subsection{Harnessing fluctuations to learn the dissipative operator}
\label{Sec:weak-GENERIC}

We here briefly describe the computational strategy of \cite{li2019harnessing} to learn a discretized version of the dissipative operator $\mathcal{K}_{z_\epsilon}$ from fluctuation data in particle simulations at a fixed value of $\epsilon$. To that end, 
the profile $z$ and the thermodynamic driving force $Q:=\delta F / \delta z$ are approximated as
$z (x, t) \approx \sum_{i} z_{i}(t) \gamma_{i}(x)$ and $ Q(x, t) \approx \sum_{i} Q_{i}(t) \gamma_{i}(x)$, where $\left\{ \gamma_i(x) \right\}$ is a set of suitable basis functions used for the discretization. Then the weak form of Eq.~\ref{Eq:T-GENERIC} can be approximated as
\begin{equation}
\label{Eq:Weak_T-GENERIC}
    \sum_{i} \left\langle \gamma_{j}, \gamma_{i} \right\rangle
    \dot{z}_i
    = -\sum_{i} \left\langle \gamma_{j}, \K_{z} \gamma_{i} \right\rangle Q_{i} 
    \quad \text{for all } j,
\end{equation}
where the bracket $\langle\cdot, \cdot\rangle$ denotes the $L^2$ inner product. Here $\left\langle \gamma_{j}, \mathcal{K}_{z} \gamma_{i} \right\rangle$ represents the discretized dissipative operator (now a finite-dimensional matrix), whose entries may be computed by the covariation of the rescaled fluctuations as
\begin{equation}
\begin{split}
\label{Eq:ComputeK}
    \left\langle\gamma_{j}, \mathcal{K}_{z} \gamma_{i}\right\rangle
    & = \lim _{h \searrow 0} \frac{1}{2 h} \mathbb{E}\left[\left(Y_{\gamma_{i}}\left(t_{0}+h\right)
    - Y_{\gamma_{i}}\left(t_{0}\right)\right) \right.
    \\
    & \quad \quad
    \left. \cdot \left(Y_{\gamma_{j}}\left(t_{0}+h\right)-Y_{\gamma_{j}}\left(t_{0}\right)\right)\right].
\end{split}
\end{equation}
Here, $t_0$ is an initial time, arbitrary as long as the system has reached a local equilibrium for the profile $z(x,t)$, $h$ is a time step that is infinitesimally small from a macroscopic perspective, yet, sufficiently large for the system to exhibit some stochastic events at the microscale, $Y_\gamma = \lim _{\epsilon \rightarrow 0} \left\langle z_{\epsilon}-z, \gamma\right\rangle / \sqrt{\epsilon}$ are the rescaled fluctuations in weak form and $z = \mathbb{E} \left[ z_{\epsilon} \right]$. In practice, multiple realizations of a particle simulation at a fixed value of $\epsilon$ and given profile $z(x)$ are performed, and the simulation domain is discretized with shape functions $\left\{ \gamma_i(x) \right\}$. The values of $\langle z_{\epsilon},\gamma \rangle$ are then recorded at times $t_0$ and $t_0+h$ to compute the right-hand side of Eq.~\ref{Eq:ComputeK}. 

To fully tabulate the discretized operator $\left\langle\gamma_{j}, \mathcal{K}_{z} \gamma_{i}\right\rangle$ and enable in such a way standalone macroscopic simulations, its input, $z(x)$, shall also be discretized. The approach we will here use to that regard differs from that of \cite{li2019harnessing}, and this will be described in further detail in Section \ref{Sec:K_Discrete}.

\section{Statistical-Physics-Informed Neural Networks (Stat-PINNs)}
\label{Sec:Stat-PINNs}

In this section we are going to introduce a machine learning framework, which we refer to as Statistical-Physics-Informed Neural Networks (Stat-PINNs), to learn the full evolution equation described by Eq.~\ref{Eq:T-GENERIC} (including both the dissipative operator $\K_z$ and the free energy $F[z]$) from particle simulations spanning a macroscopically small time step. First, in Sec.~\ref{Sec:K_Discrete}, we will introduce a structure-preserving parameterization of the discretized operator entries $\left\langle \gamma_{j}, \mathcal{K}_{z} \gamma_{i}\right\rangle$, so that its input, $z(x)$, is also discretized. Next, in Sec.~\ref{Sec:fe_Discrete}, we will discuss various common dependences of the free energy functional on $z(x)$ and  the corresponding structure for the evolution equations. Finally, we will describe the architecture of Stat-PINNs in Sec.~\ref{Sec:Stat-PINNs_Archi}, which includes two sequential neural networks for learning the dissipative operator $\K_z$ and the free energy $F[z]$.

\subsection{Structure-Preserving Parameterization of the Discretized Dissipative Operator}
\label{Sec:K_Discrete}

As previously noted in Sec.~\ref{Sec:Theory}, the dissipative operator $\K_z$ is linear as an operator acting on the thermodynamic driving force, symmetric, and positive semi-definite. Furthermore, $\K_z$ may be subjected to additional constraints when the field(s) $z$ represent conserved quantities such as mass or energy. In order to learn a thermodynamic-consistent and numerically stable structure for its discretized version from particle data, these properties should be  retained during the learning strategy. We will here discuss how to preserve such properties in the context of a conserved filed $z$ (linearity is automatically satisfied and hence omitted from the discussions). 

For simplicity, we consider a one-dimensional problem, and choose linear finite element shape functions $\{ \gamma_i(x) \}$, satisfying $\gamma_i(x_j) = \delta_{ij}$, with $x_i = i \Delta x_\gamma$ and $i=0,1,2, \cdots, N_\gamma$. This choice provides a piecewise linear approximation for both $z$ and $Q$, and is sufficient to characterize dissipative operators containing up to second-order derivatives. Notably, the local support of the shape functions can result in a sparse matrix $\left\langle\gamma_{j}, \mathcal{K}_{z} \gamma_{i}\right\rangle$, highly simplifying the learning strategy, as will be discussed next.

In general, the dissipative operator $\K_z$ can have nonlocal dependencies on both $z$ and the thermodynamic force $Q$ being acted on. Here, we classify the nonlocality of $\K_z$ into three categories. First, $\K_z$ may be local on both $z$ and $Q$, such as $\K_z Q= - \nabla \cdot \left( z(x,t) \nabla Q \right)$. In this case, $\K_z \gamma_i(x)$ has the same local support as $\gamma_i(x)$ on $x \in [x_{i-1}, x_{i+1}]$ and only depends on the profile of $z(x,t)$ within that support. Second, $\K_z$ may be local on the acting force $Q$ but nonlocal on the profile $z$, such as $\K_z Q = - \nabla \cdot \left( \int_{-L}^L z(x-\xi, t) d\xi \nabla Q \right)$. Here, $\K_z \gamma_i(x)$ still has the same local support as $\gamma_i(x)$ but depends on a wider range of the profile $z$. Third, $\K_z$ may be nonlocal on $Q$, such as $\K_z Q = \int \kappa \left( \left| x - \xi \right| \right) \left[ z(x-\xi,t) + z(\xi-x,t)\right] Q(\xi) d\xi$, where $\kappa(\cdot)$ is a kernel function. In this case, $\K_z \gamma_i(x)$ has a larger local support than $\gamma_i(x)$ and may even be non-zero on the full domain. In the following, we will focus on the first and second cases, and refer the reader to Appendix~\ref{Append:NonlocalK} for a detailed discussion on the fully nonlocal case.

When the operator $\K_z$ is local on the driving force $Q$ (first and second cases listed above), $\K_z \gamma_i(x)$ has the same local support as $\gamma_i(x)$. As a result, the discretized dissipative operator becomes a tridiagonal matrix, i.e.,  $\left\langle \gamma_{j}, \mathcal{K}_{z} \gamma_{i} \right\rangle$ non-zero only when $j \in \{i-1, i, i+1\}$. Under this assumption, which may be verified numerically using Eq.~\ref{Eq:ComputeK}, \cite{li2019harnessing} parameterized these three operator entries by the values of the local profile and its gradient at point $x_i$, i.e., $z_i$ and $\left. \nabla z \right|_i$. However, this approximation breaks the symmetry of the operator and may induce, based on our observations, numerical instabilities when solving Eq.~\ref{Eq:Weak_T-GENERIC}. To preserve the operator symmetry, we here regard the entries of the tridiagonal matrix as two functions of several discrete values of $z$. Specifically, 
\begin{equation}
\begin{split}
\label{Eq:K0_K1_Z}
    & K_0(\mathbf{Z}_{i}^0)
    = \left\langle \gamma_{i}, \mathcal{K}_{z} \gamma_{i} \right\rangle ,
    \quad \text{and} \\
    & K_1(\mathbf{Z}_{i}^1)
    = \left\langle \gamma_{i+1}, \mathcal{K}_{z} \gamma_{i} \right\rangle 
    = \left\langle \gamma_{i}, \mathcal{K}_{z} \gamma_{i+1} \right\rangle,
\end{split}
\end{equation}
where $\mathbf{Z}_{i}^0 = (z_{i-1}, z_i, z_{i+1})$ and $\mathbf{Z}_{i}^1 = (z_i, z_{i+1})$ when the operator is local on $z$ (or the nonlocal effect is negligible within the given numerical discretization).
If $\K_z$ is nonlocal on $z$, then additional neighboring points may be needed, e.g., $\mathbf{Z}^0_i = (z_{i-2}, z_{i-1}, z_i, z_{i+1}, z_{i+2})$ and $\mathbf{Z}^1_i = (z_{i-1}, z_i, z_{i+1}, z_{i+2})$.
We remark that this multi-point expansion method on operator entry $K_p$  with $p=0, 1$ can be equivalently written as $K_p(\mathbf{Z}^p_i) = \tilde{K}_p(\tilde{\mathbf{Z}}^p_{i+\frac{p}{2}})$ to change the functional dependency from profiles at multiple local points $\mathbf{Z}^p_i$ to the value of the profile and its derivatives $\tilde{\mathbf{Z}}^p_{i+\frac{p}{2}}$ at the middle point $x_{i+\frac{p}{2}} = (x_{i} + x_{i+p})/2$ (see Appendix~\ref{Append:MP&MidP} for further details). 

Next, we turn our attention to the constraints on the operator when the field $z$ obeys a conservation law of the form $\frac{d}{dt}\int z(x) \, dx =0$, such as the density in the diffusive system that will be discussed in Sec.~\ref{Sec:Arrhenius}. In this case, the operator entries should satisfy $\sum_{j} \left\langle \gamma_{j}, \K_{z} \gamma_{i} \right\rangle = 0$ \citep{li2019harnessing}, and, thus
\begin{equation}
\label{Eq:K0_MassCons}
     K_0(\mathbf{Z}^0_i)
     = - K_1(\mathbf{Z}^1_{i-1}) 
     - K_1(\mathbf{Z}^1_i).
\end{equation}
That is, the whole discretized operator can be written as a function of the off-diagonal entry $K_1$.

Finally, the discretized operator should be positive semi-definite for an arbitrary profile $z(x)$. Although the Cholesky decomposition is often used to guarantee that a given matrix is positive semi-definite, we here use an alternative approach, which will prove to be much simpler to implement. We first note that, by combining Eqs.~\ref{Eq:Weak_T-GENERIC} and \ref{Eq:K0_MassCons}, the evolution equation of the system can be re-written as
\begin{equation}
\begin{split}
\label{Eq:Weak_T-GENERIC_MassCons}
    \sum_{i=j-1}^{j+1} 
    \left\langle \gamma_{j}, \gamma_{i} \right\rangle
    \frac{\partial z_i}{\partial t} 
    & = -
    \Delta_b \left[  K_1(\mathbf{Z}^1_j)
    \Delta_f Q_j \right],
\end{split}
\end{equation}
where $\Delta_f B_i = B_{i+1} - B_i$ is the forward finite difference for an arbitrary field $B$ and $\Delta_b B_i = B_{i} - B_{i-1}$ is the backward finite difference. Equation~\ref{Eq:Weak_T-GENERIC_MassCons} therefore suggests that $-\K_z$ is a second-order diffusion type operator with $-K_1$ acting as the mobility coefficient. 
As rigorously shown in Appendix~\ref{Append:NonlocalK}, the discretized operator can then be enforced to be positive semi-definite by simply requiring $K_1 \leq 0$ for an arbitrary profile.

\subsection{Free Energy Parameterization and Ensuing Equation Structure}
\label{Sec:fe_Discrete}

The thermodynamic driving force for systems governed by Eq.~\ref{Eq:T-GENERIC} is given by the functional derivative of the free energy, i.e., $Q = \delta F/\delta z$. We will here  assume, as it is common in continuum descriptions, that the free energy functional has a density $f$ associated to it, so that $F[z] = \int f dx$. Yet, we will allow for various dependencies of $f$ on the profile $z$, which  will result in different thermodynamic forces and, consequently, in different types of evolution equations. 

The simplest case is that of $f=f(z)$, where the free energy density is only a function of the local profile, and the thermodynamic force is then given by
\begin{equation}
\label{Eq:Q_f(z)}
    Q = f'(z).
\end{equation}
For a discretized dissipative operator of the form of Eq.~\ref{Eq:Weak_T-GENERIC_MassCons}, the evolution equation will then be a standard second-order diffusion equation.

\begin{figure*}
\centering
\includegraphics[width=0.8\linewidth]{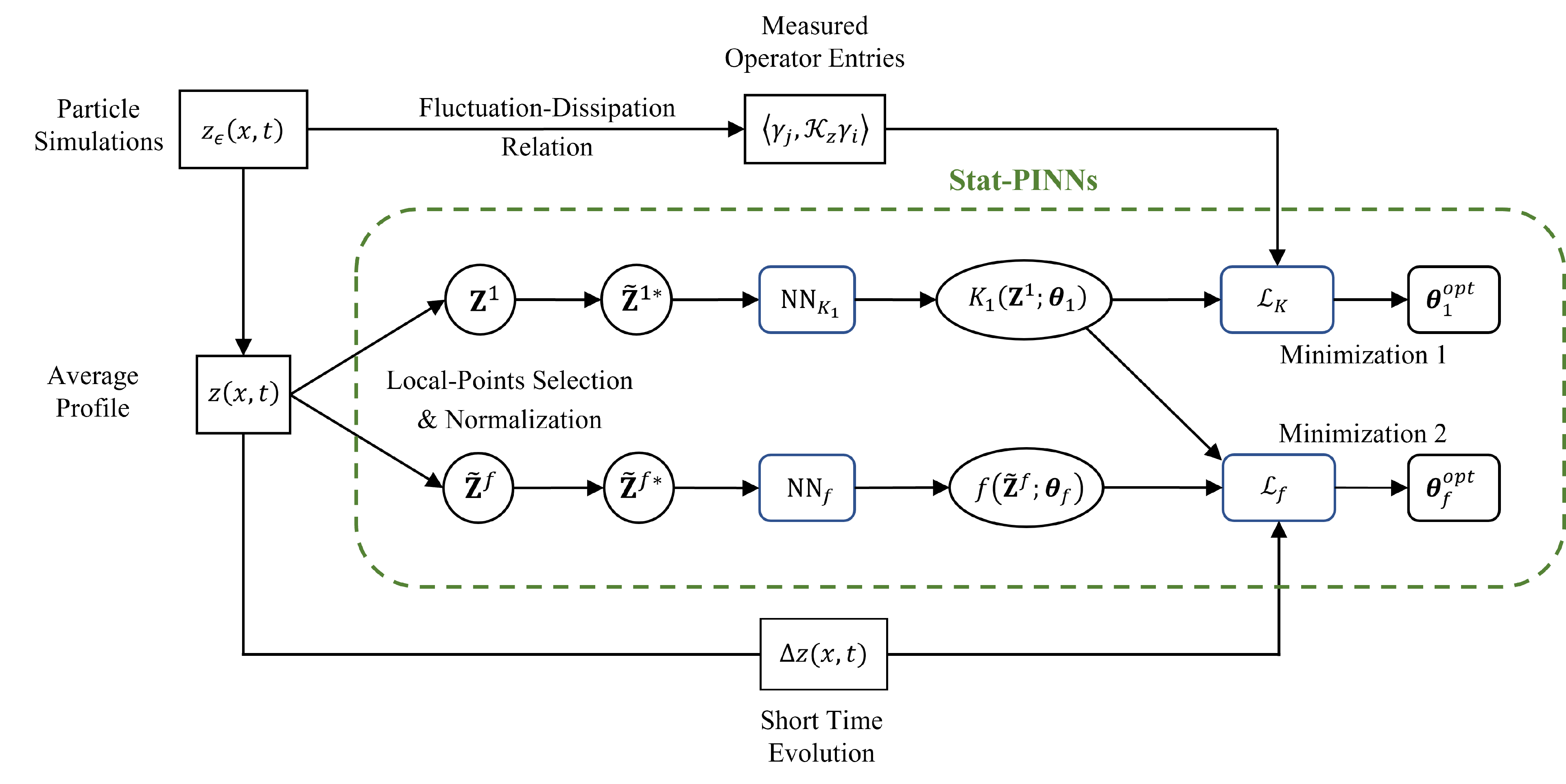}
\caption{Architectures for Stat-PINNs.}
\label{fig:Stat-PINNs}
\end{figure*}

If the free energy density also depends on the local gradient, i.e., $f = f(z, \nabla z)$, the thermodynamic force is then given by
\begin{equation}
\label{Eq:Q_f(z,grad_z)}
    Q = \frac{\partial f}{\partial z} 
     - \nabla \cdot \frac{\partial f}{\partial \nabla z}.
\end{equation}
We remark that if we use a first-order backward (or forward) finite difference scheme to discretize $\nabla z$, the divergence in Eq.~\ref{Eq:Q_f(z,grad_z)} should be expressed as a forward (or backward) scheme, respectively (see Appendix~\ref{Append:fe_Q} for further details). In the discrete setting, $f$ can be viewed as a function of two local points and $Q$ as a function of three local points. Following the operator form given in Eq.~\ref{Eq:Weak_T-GENERIC_MassCons}, the evolution equation is now a fourth-order equation, such as a Cahn–Hilliard equation. 

The free energy density may also have a dependency on higher order derivatives. For instance, if it also depends on the Laplacian (denoted as $\nabla^2 z$ for simplicity for a  one-dimensional system), i.e., $ f = f(z, \nabla z, \nabla^2 z)$, the thermodynamic force now includes an additional term,
\begin{equation}
\label{Eq:Q_f(z,grad_z,lap_z)}
    Q = \frac{\partial f}{\partial z} 
      - \nabla \cdot \frac{\partial f}{\partial \nabla z}
      + \nabla^2 \frac{\partial f}{\partial \nabla^2 z}.
\end{equation}
In the discrete setting, if we use a backward, forward or central scheme to calculate $\nabla z$ and use a second-order central scheme for $\nabla^2 z$, the discretized $f$ can be viewed as a function of three local points and $Q$ as a function of five local points (see Appendix~\ref{Append:fe_Q} for more details). Equation~\ref{Eq:Weak_T-GENERIC_MassCons} is now a sixth-order partial differential equation.

The above discussion highlights how higher-order terms in the dependence of the free energy density increase the order of the resulting PDE. Yet, it is interesting to note that, regardless of the order of the PDE (it being fourth- or even sixth- order), one can still use linear shape functions to discretize the dissipative operator, as this only contains second order derivatives under the assumptions made in Section \ref{Sec:K_Discrete}. Testing examples can be found in \ref{App:Cahn-Hilliard}.

\subsection{Stat-PINNs Architecture}
\label{Sec:Stat-PINNs_Archi}


We will now introduce a machine learning architecture, which we denote as Statistical-Physics-Informed Neural Networks (Stat-PINNs), to learn the coarse-grained dissipative evolution equation. This architecture is depicted in Fig.~\ref{fig:Stat-PINNs} and consists of two neural networks that will be trained sequentially. These neural networks are aimed at learning the discretized operator $\K_z$ and the free energy density $f$, based on the parameterizations discussed in Sections \ref{Sec:K_Discrete} and \ref{Sec:fe_Discrete}, respectively. 

The first neural network is constructed to learn the off-diagonal entry $K_1$, as this is sufficient to fully characterize the discretized operator (see Section \ref{Sec:K_Discrete}). Such an entry will be learned from multiple realizations of short-time particle simulations using Eq.~\ref{Eq:ComputeK} and Eq.~\ref{Eq:K0_MassCons} so that both the diagonal and off-diagonal entries are well-approximated. 
We denote the measured data of the diagonal operator entries, off-diagonal operator entries and corresponding field values as 
$\left\{K_0^{(s)}, K_1^{(s)}, \mathbf{Z}_{i^s}^{0(s)} \right\}_{s=1}^{N_K}$, 
where $N_K$ is the number of training data and $i^s$ denotes the spatial index for the $s$-th sample. We remark that $\mathbf{Z}_{i^s}^{0(s)}$ includes both $\mathbf{Z}_{i^s-1}^{1(s)}$ and $\mathbf{Z}_{i^s}^{1(s)}$, which are the field values needed in the loss function, as will be seen next.
The off-diagonal entry $K_1(\mathbf{Z}^1; \boldsymbol{\theta}_1)$ is represented through a neural network $\text{NN}_{K_1}(\tilde{\mathbf{Z}}^{1*}; \pmb{\theta}_1)$, applied onto a non-positive function $g(\cdot)$, i.e., $g(\text{NN}_{K_1})$, to ensure the positive semi-definiteness of the operator. Here, $\tilde{\mathbf{Z}}^{1*}$ is the normalized version of $\mathbf{Z}^1=\left\{\mathbf{Z}_{i^s}^{1(s)} \right\}_{s=1}^{N_K}$ and $\theta_1$ represents the trainable parameters. Further details on the relation between $K_1$ and $\text{NN}_{K_1}$ can be found in Appendix~\ref{Append:Stat-PINNs_Details}.
Then, the training is performed by minimizing the following loss function
\begin{equation}
\begin{split}
\label{Eq:Loss_K_MassCons}
    & \mathcal{L}_K 
    = 
    \frac{\lambda_1}{2N_1} 
    \sum_{s=1}^{N_1} \left\|
    K_{1} \left( \mathbf{Z}_{i^s}^{1(s)}; \pmb{\theta}_1 \right)
    - K_1^{(s)} \right\|^2
    \\
    & +
    \frac{\lambda_0}{2N_0} 
    \sum_{s=1}^{N_0} \left\| 
    - K_{1} \left( \mathbf{Z}_{i^s-1}^{1(s)}; \pmb{\theta}_1 \right)
    - K_{1} \left(\mathbf{Z}_{i^s}^{1(s)}; \pmb{\theta}_1 \right)
    - K_0^{(s)} \right\|^2,
\end{split}
\end{equation}
where $\lambda_0$ and $\lambda_1$ are two adaptive loss weights determined automatically based on the neural tangent kernel method \citep{wang2022and}.

Once the dissipative operator $\K_z$ is learned from the first neural network, we can learn the free energy density $f(\tilde{\mathbf{Z}}^f)$ from macroscopic evolutions over a small time step $\Delta t$. We approximate $f$ through a second neural network $\text{NN}_f(\tilde{\mathbf{Z}}^{f*}; \boldsymbol{\theta}_f)$, where $\boldsymbol{\theta}_f$ denotes the trainable parameters for $\text{NN}_f$, $\tilde{\mathbf{Z}}^f$ denotes the value of the local field and potentially its spatial derivative, and $\tilde{\mathbf{Z}}^{f*}$ is the corresponding normalized version (see Appendix~\ref{Append:Stat-PINNs_Details} for details). 
The training data is denoted as $\left\{ \mathbf{Z}^{Q(s)}_{i^s}, \Delta \mathbf{Z}^{Q(s)}_{i^s} \right\}_{s=1}^{N_f}$, where $\mathbf{Z}^{Q(s)}_{i^s}$ represents the local field values needed for computing the thermodynamic force $Q(\mathbf{Z}^{Q(s)}_{i^s}; \boldsymbol{\theta}_f)$ at point $x_{i^s}$ through $\text{NN}_f$, and $\Delta \mathbf{Z}^{Q(s)}_{i^s}$ represents the time increments of these field values during the time interval $\Delta t$.
We remark that both the local field values and their time increments are averaged from $R$ realizations of macroscopically equivalent particle simulation. 
Here, the time interval $\Delta t$ should be macroscopically small, but still much longer than the infinitesimal timestep $h$ used for calculating the operator entries. Since the deterministic increment in Eq.~\ref{Eq:Sto_T-GENERIC} is proportional to $\Delta t$ and the stochastic noise is proportional to $\sqrt{\Delta t}$, $\Delta z$ will in general be noisy. We should therefore construct the loss function not only based on the deterministic Eq.~\ref{Eq:Weak_T-GENERIC} as Physics-Informed Neural Networks (PINNs), but also taking into account the effect of the stochastic noise term as \citep{dietrich2023learning}
\begin{equation}
\begin{split}
\label{Eq:Loss_feNN}
    \mathcal{L}_f
    & = \frac{1}{2N_f} \sum_{s=1}^{N_f} 
    \frac{1}{ \left( \sigma_{Eq,j^s}^{(s)} \right)^2}
    \left\| \sum_{i} \langle \gamma_{j^s}, \gamma_{i} \rangle
    \frac{\Delta z_i^{(s)}}{\Delta t}
    \right.
    \\
    & \quad
    \left.
    + \sum_{i} \langle \gamma_{j^s}, \K_{z^{(s)}} \gamma_i \rangle 
    Q(\mathbf{Z}^{Q(s)}_{i}; \boldsymbol{\theta}_f)
    \right\|^2
.
\end{split}
\end{equation}
Here, $ \langle \gamma_{j^s}, \K_{z^{(s)}} \gamma_i \rangle $ is calculated from the trained neural network $\text{NN}_{K_1}$ and the local field values, the thermodynamic forces $Q_i$ are parameterized by the neural network $\text{NN}_f$ according to the discussion in Sec.~\ref{Sec:fe_Discrete},
and $\sigma_{Eq,j}^2$ represents the variance of the stochastic noise at point $x_j$, which is equal to $ 2 \epsilon \langle \gamma_{j^s}, \K_{z^{(s)}} \gamma_{j^s} \rangle / (R \Delta t)$ (see Appendix~\ref{Append:Var_Weak_Eq} for further details).
This denominator can normalize the noise at different points  to a standard Gaussian distribution and is crucial for the training process. This is particularly the case when the noise is comparable to or dominant over the deterministic term and when the variance of noise varies among different data points.

\section{Example: Arrhenius Interacting Particles Processes}
\label{Sec:Arrhenius}

\subsection{Particle Process and Existing Macroscopic Model for Arrhenius Diffusion}
\label{Sec:Arr_Model}

Arrhenius-type dynamics are characterized by an exponential dependence on an activation energy $\hat{E}_d$ and inverse temperature $\beta = 1/( k_B T) $, i.e., $e^{-\beta \hat{E}_d}$, and arise in a wide range of phenomena, such as surface absorption, chemical reactions, or vacancies and intersticials in solids \citep{gibbs1972sufficient,linderoth1997surface,gilmer1972simulation,laidler1984development,zhdanov1991arrhenius,dewey1994arrhenius}. 

As an illustrative example of such type of dynamics, we here consider a stochastic jumping process for interacting particles on a one-dimensional lattice \citep{vlachos2000derivation,katsoulakis2003coarse}, where
each lattice site can only be occupied by at most one particle, i.e., the occupation number is $\eta(x)=1$ if site $x$ is occupied and $\eta(x)=0$ if it is empty. The probability of a particle jumping event from site $x$ to its nearest-neighbor $y$ (left or right in 1D) is of Arrhenius type and given by $p(x \rightarrow y)= d\, \eta(x) \left( 1-\eta(y) \right) e^{-\beta \hat{U}(x)}$. Here, $d$ is the jumping frequency, and $\hat{U}(x) = \hat{U}_0 + \sum_{\xi\neq x} \hat{J}(x-\xi)\eta(\xi)$, where  $\hat{U}_0$ is the binding energy, and $\hat{J}(\cdot)$ describes the interaction energy between particles, which is attractive for $\hat{J}>0$ and repulsive for $\hat{J}<0$. Despite the apparent simplicity of this Arrhenius diffusion model, its analytic macroscopic description only exists for limited cases. Particularly, when the interaction between the particles is long range (compared to the lattice size), the macroscopic evolution for the particle density $\rho$ is given by \citep{vlachos2000derivation} 
\begin{equation}
\label{Eq:ContEq_Arr_LongL}
    \frac{\partial\rho}{\partial t}
    =\nabla \cdot \left( m\left[\rho\right]\nabla \frac{\delta F[\rho]}{\delta \rho}\right) \quad \text{and}
    \quad m\left[\rho\right] = D \rho (1-\rho) e^{- J \ast \rho},
\end{equation}
where $D=d\, e^{-\beta \hat{U}_0}$, $J(\cdot) = \beta \hat{J}(\cdot)$ is the dimensionless interaction energy, and $J\ast \rho=\int J(x-\xi)\rho(\xi)\mbox{d}\xi$
denotes the convolution between $J$ and $\rho$. Furthermore, $F$ is the (dimensionless) Helmholtz free energy given by
\begin{equation} 
\begin{split}
\label{Eq:FreeEnergy_Arr}
    F[\rho]
    & =-\int \frac{1}{2}\rho \left(J \ast \rho \right) \mbox{d}x 
    \\
    & \quad
    + \int\left[\rho\ln \rho+(1-\rho)\ln{(1-\rho)}\right]\mbox{d}x,
\end{split}
\end{equation}
which consists of a nonlocal energy term and a cross entropy term. When the interaction potential is symmetric, i.e., $J(x)=J(-x)$, the thermodynamic driving force can be written as,
\begin{equation}
\label{Eq:dFdrho}
    Q = \frac{\delta F}{\delta \rho} 
    = -J \ast \rho 
    - \ln{\left(\frac{1}{\rho}-1\right)}.
\end{equation}
Moreover, the fluctuations of the particle jumping process can be accounted for by adding the stochastic noise term $\nabla \cdot\left( \sqrt{2 \epsilon m\left[\rho\right]} \dot{W}_{x,t} \right)$ to Eq.~\ref{Eq:ContEq_Arr_LongL} \citep{vlachos2000derivation}. That is, the macroscopic description and its stochastic counterpart are of the form of Eq.~\ref{Eq:T-GENERIC} and Eq.~\ref{Eq:Sto_T-GENERIC}, respectively, with the following dissipative operator
\begin{equation}
\label{Eq:ArrK}
    \K_\rho
    =-\nabla \cdot \left( m\left[\rho\right] \nabla \right)
    = -\nabla \cdot \left[ D \rho(1-\rho)e^{- J\ast \rho} \nabla \right].
\end{equation}
Further details for the weak form of this long-range model can be found in Appendix~\ref{Append:Arr_Ana}.

\begin{figure*}
\centering
\includegraphics[width=0.85\linewidth]{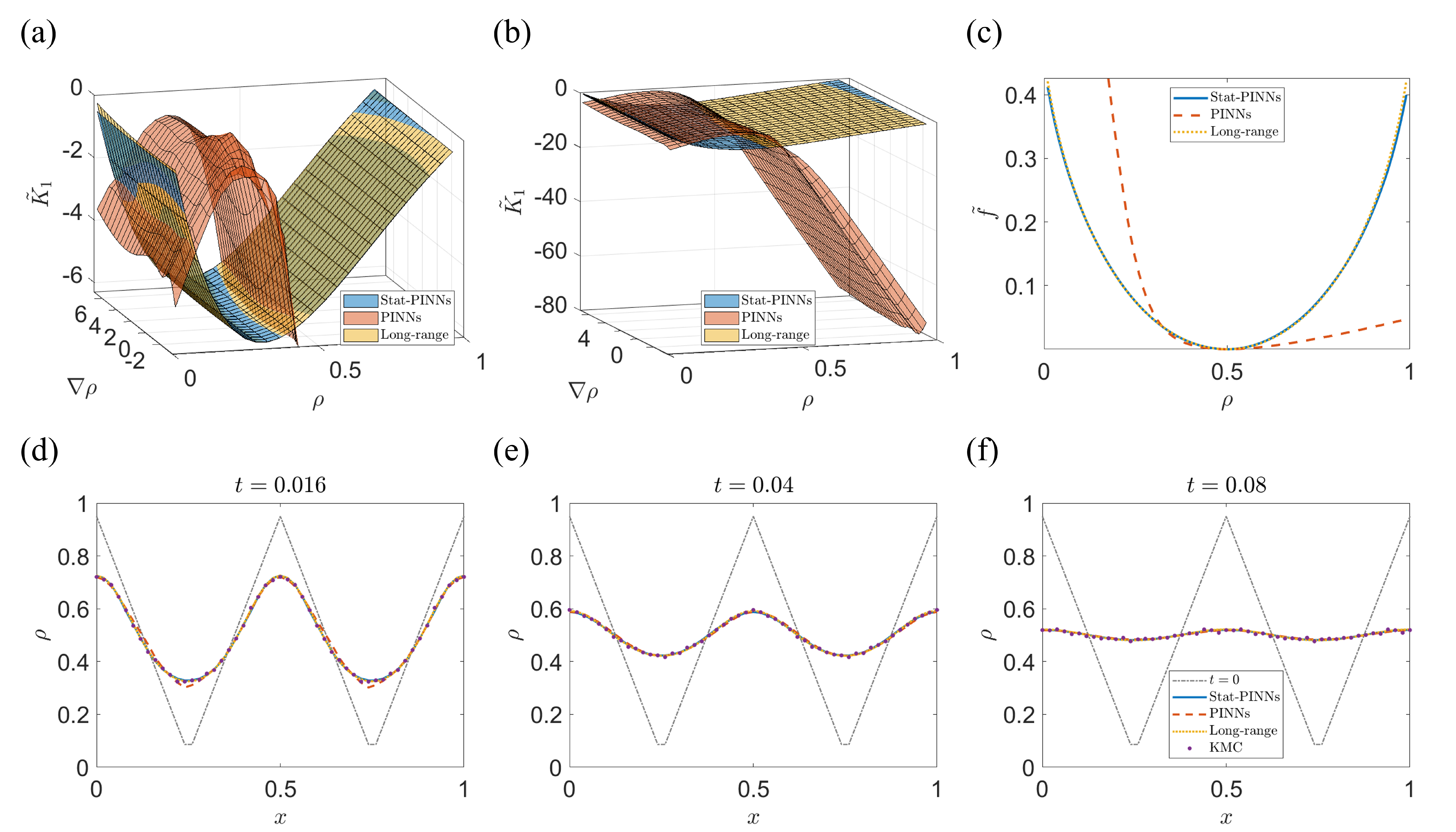}
\caption{Results for the Arrhenius process with $JL=0.9, L=40\epsilon$, including (a, b) the dissipative operator entry $\tilde{K}_1$ plotted in two different ranges, (c) the calibrated free energy density $\tilde{f}$, and (d-f) snapshots of the macroscopic evolution starting from a triangular wave initial profile (black dash dotted line). Predictions from Stat-PINNs (blue surfaces or blue solid lines), PINNs (orange surfaces or orange dashed lines) and long-range analytic model (yellow surfaces or yellow dotted lines) are shown. Results from KMC particle simulations (purple dots) are used as the true macroscopic evolution for comparison.}
\label{fig:JL0.9_L40}
\end{figure*}

A specific case of the above dynamics is that of no interaction between the particles, i.e., $\hat{U}(x)=0$. Such particle dynamics is known as the symmetric simple exclusion process (SSEP) \citep{embacher2018computing,adams2013large,huang2020particle}. Even though this ``interaction'' is not long-range, its coarse-grained evolution (both the PDE and its stochastic version) can also be given by the equations above with $J=0$ and $\tilde{U}_0 = 0$.

Beyond this long-range model described by Eqs.~\ref{Eq:ContEq_Arr_LongL}-\ref{Eq:FreeEnergy_Arr}, our understanding of the macroscopic evolution equations for the case with short-range interactions is still limited. The coarse-grained free energy expression has only been obtained from statistical mechanics for the case of nearest-neighbour interactions \citep{katsoulakis2003coarse}, whilst the analytical expression for the dissipative operator, or the full dynamics, for such nearest-neighbour interaction case is still unknown. We will thus aim to capture the full dynamics of the Arrhenius process with different interactions by the proposed Stat-PINNs framework in the following section.

\subsection{Results}
\label{Sec:Results}

We consider the Arrhenius particle process just described with $d=1$, $U_0=0$ and a symmetric step function interaction defined as $J(x) = J_0$ for $0 < |x| \leq L$ and $J(x)=0$ otherwise. 
Here, we will test our Stat-PINNs framework for three different cases of the Arrhenius diffusion process. The first one is chosen to have long-range interactions between the particles, so as to enable a direct comparison with the known analytic solution. The other two cases will correspond to weak and strong short-range interactions, and are aimed at testing the capability of the method to discover their continuum thermodynamic model, previously unknown, to the best of the authors' knowledge. In all cases, the results will be directly compared to the results from traditional PINNs, where no recourse is made to statistical mechanics. To perform a fair comparison, the same thermodynamic structure and parameterization will be used. Specifically, two neural networks for $\K_1$ and $f$ will be jointly trained using the loss function given by Eq.~\ref{Eq:Loss_feNN}, but without normalizing the residuals with the equation variance (i.e., setting the denominator $\sigma_{Eq}^2=1$).

\begin{figure*}
\centering
\includegraphics[width=0.85\linewidth]{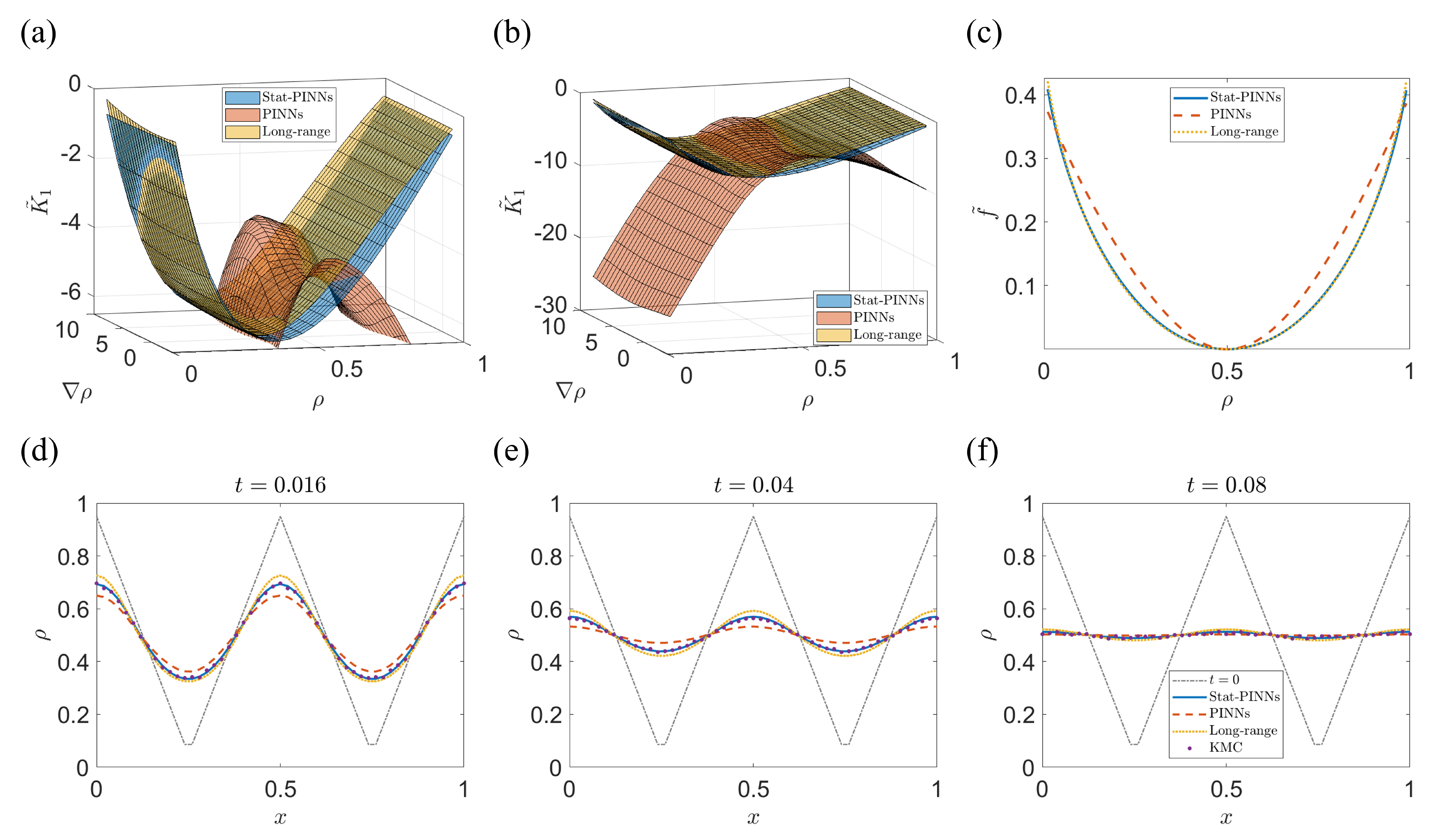}
\caption{Results for Arrhenius process with $JL=0.9, L=2\epsilon$, including (a, b) dissipative operator entry $\tilde{K}_1$ plotted in two different ranges, (c) calibrated free energy density $\tilde{f}$, and (d-f) snapshots of macroscopic evolution starting from a triangular wave initial profile (black dash dotted line). Predicting methods include Stat-PINNs (blue surfaces or blue solid lines), PINNs (orange surfaces or orange dashed lines) and long-range analytic model (yellow surfaces or yellow dotted lines). Results from KMC particle simulation (purple dots) are used as true macroscopic evolution for comparison.}
\label{fig:JL0.9_L2}
\end{figure*}

For each case, the data is collected from a one-dimensional particle system with $N_b = 2000$ bins and 25 shape functions using KMC particle simulations starting from 28 different cosine initial profiles. This particle system is then rescaled to the macroscopic range $x \in [0,0.5]$, indicating a lattice size $\epsilon = 2.5\times 10^{-4}$ and a spatial discretization $\Delta x_\gamma = 0.02$. The restriction to the domain $[0,0.5]$ is motivated by symmetry considerations to facilitate the particle simulations used for comparison, see Appendix~\ref{Append:KMC_Settings}. Each data point for the operator entries is calculated from 10 sequential $h$ intervals in $R=10^4$ realizations, i.e., expectations are approximated by performing averages over $R_K = 10^5$ realizations for each initial profile, where the small time interval $h$ is taken as $h=0.01 \epsilon^2 \tau$, with $\tau=\exp\left(2 J_0 L \rho_{\max}\right)$ a factor that depends on the maximum density $\rho_{\max}$ of each simulated profile. Each data point for learning the free energy is generated from these same $R=10^4$ realizations for each initial profile, within a short time interval $\Delta t >> h$, chosen to be about $20\epsilon^2 \tau \sim 40\epsilon^2 \tau$ depending on each case. More details about the particle simulations and corresponding parameter settings can be found in Appendix~\ref{Append:KMC_Settings}. 
Training details are included in Appendix~\ref{Append:Training_Settings}. Using the learnt operator and free energy, Eq.~\ref{Eq:Weak_T-GENERIC} is then solved using a forward Euler scheme and the results are compared to those stemming from the long-range analytical model (Eqs.~\ref{Eq:ContEq_Arr_LongL} and \ref{Eq:FreeEnergy_Arr}) together with the average of 600 realizations from direct KMC simulations using the Bortz–Kalos–Lebowitz (BKL) algorithm \citep{bortz1975new}. The macroscopic predictions are solved on $x\in[0,1]$ with spatial discretization $\Delta x_\gamma=0.02$ and periodic boundary conditions. The initial condition chosen to test the predictive capability of Stat-PINNs is a triangular profile, and hence distinct from the training profiles. We remark that while the training profiles are simulated only for a $\Delta t$ time interval, the macroscopic simulations can be performed over an arbitrary large time interval, as well as arbitrary initial conditions, without needing to perform extrapolations, as long as the macroscopic evolution lies within the trained phase space, e.g., range of $\rho$ and $\nabla \rho$ explored within the profiles used in the training process.

All KMC particle simulations are coded in C++ and run on Intel(R) Xeon(R) CPUs E5-2683 v4 @ 2.10 GHz.
Stat-PINNs and the corresponding continuum predictions are coded in Python. JAX is the main Python library for implementing neural networks. Other standard libraries such as Numpy, JAX Numpy, Scipy and Matplotlib are used for data pre- and post-processing.


\begin{figure*}
\centering
\includegraphics[width=0.85\linewidth]{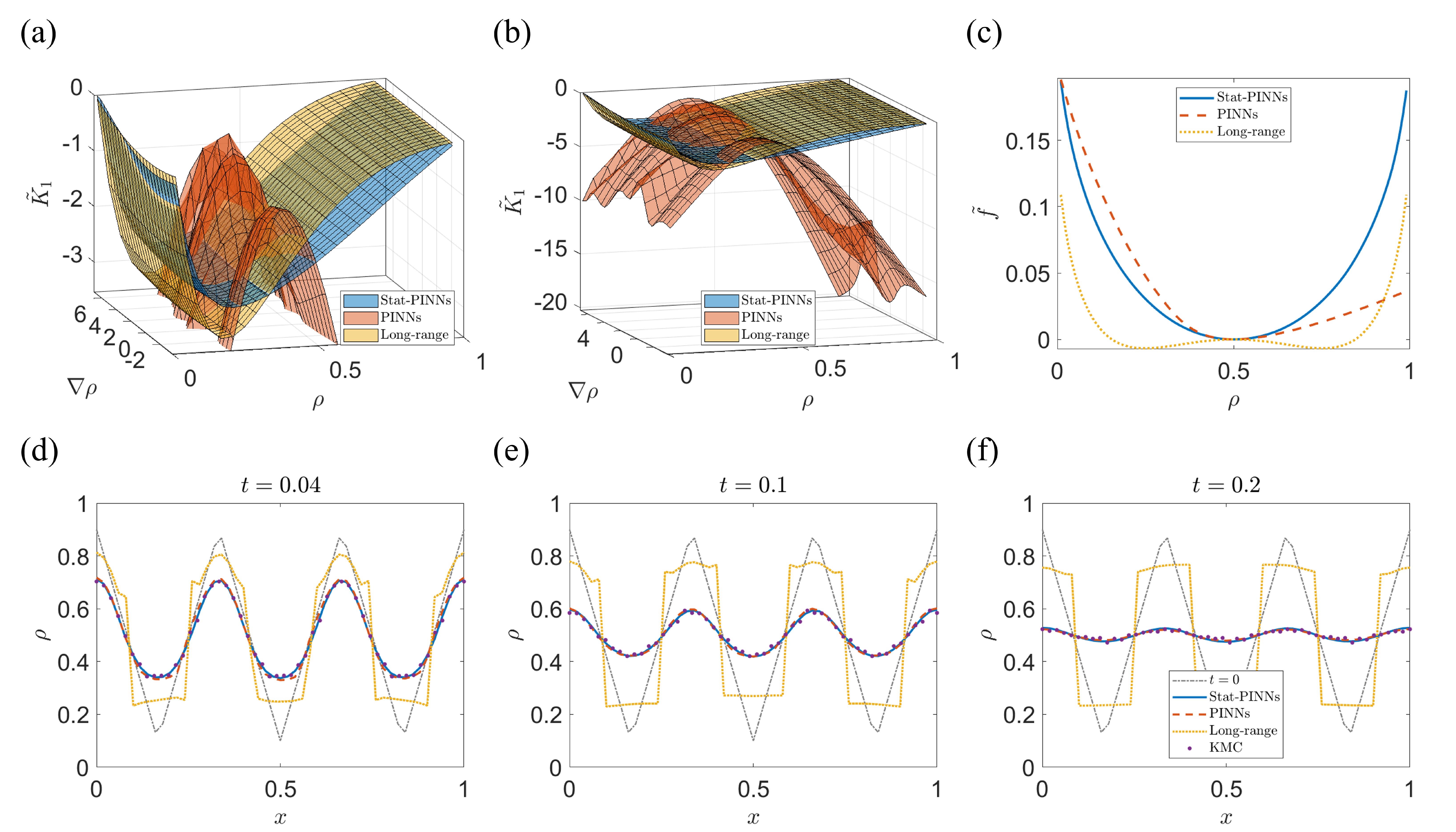}
\caption{Results for Arrhenius process with $JL=2.2, L=2\epsilon$, including (a, b) dissipative operator entry $\tilde{K}_1$ plotted in two different ranges, (c) calibrated free energy density $\tilde{f}$, and (d-f) snapshots of macroscopic evolution starting from a triangular wave initial profile (black dash dotted line). Predicting methods include Stat-PINNs (blue surfaces or blue solid lines), PINNs (orange surfaces or orange dashed lines) and long-range analytic model (yellow surfaces or yellow dotted lines). Results from KMC particle simulation (purple dots) are used as true macroscopic evolution for comparison.}
\label{fig:JL2.2_L2}
\end{figure*}

Figure~\ref{fig:JL0.9_L40} shows the coarse-grained results from a particle system with long-range interactions with $J_0 L = 0.9$ and $L = 40 \epsilon$, i.e., each particle interacts with other particles within 40 neighboring lattice sites on each side. Given that the interaction is weak, we regard the dynamics as local within numerical accuracy even though the interaction range $L=\Delta x_\gamma / 2$ is long from a particle perspective. We set the off-diagonal entry $K_1$ as a function of two local points, or equivalently as $\tilde{K}_1(\rho, \nabla \rho)$, with $\rho$ and $\nabla \rho$ evaluated at the middle point of the support, and the free energy density as $f(\rho)$. Figures~\ref{fig:JL0.9_L40}a-\ref{fig:JL0.9_L40}c show the comparison between the Stat-PINNs result (blue) and long-range analytic model (yellow) for the off-diagonal entry $\tilde{K}_1$, shown for two different ranges, and the free energy density.  The latter is calibrated by setting its value and that of its derivative to zero at $\rho=0.5$ in order to cancel the non-uniqueness due to the structure of the operator and the thermodynamic force $Q$  (see Appendix~\ref{Append:fe_non-unique} for details), and it is denoted as $\tilde{f}(\rho)$ after calibration. We see that the prediction of Stat-PINNs (blue) has an excellent agreement with the long-range analytic model (yellow), exhibiting a relative $L^2$ error of $0.59\%$  for $\tilde{K}_1$ and $2.56\%$ for $\tilde{f}$. In contrast, PINNs (orange) cannot uniquely identify the true operator and free energy. Furthermore, the operator entry identified by PINNs differs in magnitude for low and high densities, exhibits an opposite trend, and is non-smooth. The learned free energy from PINNs even lacks symmetry. The learned operator and free energy from both Stat-PINNs and PINNs are used to solve Eq.~\ref{Eq:Weak_T-GENERIC} for a triangular wave initial profile with average density $\rho_{ave} = 0.5$, amplitude $A=0.45$ and frequency $f=2$, and the results are compared to that of the analytical long-range evolution equation as well as direct particle simulations. The total time simulated is $t_{tot}=0.08$, and the timestep used in the discretization is $\Delta t_{sim} = \Delta x_\gamma^2/5 = 8 \times 10^{-5}$. Figures~\ref{fig:JL0.9_L40}d-\ref{fig:JL0.9_L40}f show the results for the macroscopic evolution. Both the prediction from Stat-PINNs (blue solid line) and the solution from the long-range analytic model (yellow dotted line) agree with the KMC particle simulation (purple dots)
In contrast, there exists a noticeable error in the prediction from PINNs (orange dashed line) at the beginning of the evolution.
These results therefore show that Stat-PINNs can not only predict the correct coarse-grained macroscopic evolution, but also learn the true thermodynamics and kinetics of the system.


Figure~\ref{fig:JL0.9_L2} shows the results from a short-range weakly interacting Arrhenius process with $J_0 L = 0.9$ and $L = 2 \epsilon$, where the long-range analytic model is not expected to work and no analytic solution is known, to the best of the authors' knowledge. For this case, we also consider a local dependence of the operator and free energy density of the form $\tilde{K}_1(\rho, \nabla \rho)$ and  $f(\rho)$. Figures~\ref{fig:JL0.9_L2}a-\ref{fig:JL0.9_L2}c show the comparison between the results from PINNs (orange), Stat-PINNs (blue) and the long-range analytical model (yellow) for the off-diagonal entry $\tilde{K}_1$, shown for two different ranges, and the calibrated free energy density $\tilde{f}(\rho)$, respectively. While the predicted free energy from Stat-PINNs agrees with Eq.~\ref{Eq:FreeEnergy_Arr} with a $1.85\%$  relative $L^2$ error, the predicted operator is different from the long-range model as could be expected (there is a $13.83\%$ $L^2$ relative difference). Predictions for the macroscopic evolution of a triangular wave initial profile with average density $\rho_{ave} = 0.5$, amplitude $A=0.45$ and frequency $f=2$ are shown in Figs.~\ref{fig:JL0.9_L2}d-\ref{fig:JL0.9_L2}f. These are solved for a total time $t_{tot} = 0.08$ using a timestep $\Delta t_{sim} = \Delta x_\gamma^2 / 5 = 8 \times 10^{-5}$.  While  the Stat-PINNs prediction perfectly agrees with the true evolution from KMC simulations, the results from PINNs exhibit a noticeable error and so does as well the long-range analytical model (particularly in the higher density range). In particular, the long-range anayltical prediction appears to lag behind the true evolution (which coincides with that of Stat-PINNs). This difference can be explained from the predicted operator entry shown in Fig.~\ref{fig:JL0.9_L2}a, which can be physically interpreted as a mobility coefficient. The value of $K_1$ predicted with Stat-PINNs has a larger absolute value compared to the long-range model, and this difference is more significant in the higher density range. 


\begin{table*}
\centering
\begin{tabular}{|c | c| c | c | c|} 
 \hline
 Case 
 &  \makecell{Stat-PINNs \\ Data Collection}
 &  \makecell{Stat-PINNs \\Training}
 &  \makecell{Stat-PINNs \\Prediction}
 &  \makecell{KMC for\\ Full Model}
 \\
 \hline
 \makecell{$J_0 L = 0.9$\\$L=40\epsilon$}
 & 55 days ($\div 28 \div 10^{4}$)
 & $\sim20$ min
 & $<5$ min
 & 3500 days ($\div 600$)
 \\
 \hline
 \makecell{$J_0 L = 0.9$\\$L=2\epsilon$} 
 & 19 days ($\div 28 \div 10^{4}$)
 & $\sim20$ min
 & $<5$ min
 & 875 days ($\div 600$)
 \\
 \hline
 \makecell{$J_0 L = 2.2$\\$L=2\epsilon$}
 & 39 days ($\div 28 \div 10^{4}$)
 & $\sim20$ min
 & $<5$ min
 & 625 days ($\div 600$)
 \\
 \hline
\end{tabular}
\caption{Computational costs for Stat-PINNs compared with KMC particle simulations. Here the numbers with units of days  represent the total cost for all realizations as if they were perform in the absence of any parallelization and the number in the following parenthesis denotes the maximum factor for increasing computational efficiency by parallelizing all realizations.}
\label{Table:Computation_Cost}
\end{table*}

Figure~\ref{fig:JL2.2_L2} shows the results from a short-range strongly interacting Arrhenius particle process with $J_0 L = 2.2$ and $L = 2 \epsilon$. Here, we also assume that the off-diagonal entry $K_1$ and the free energy density are local and of the form $\tilde{K}_1(\rho, \nabla \rho)$ and $f(\rho)$, respectively. As shown in Figs.~\ref{fig:JL2.2_L2}a-\ref{fig:JL2.2_L2}c, the predictions for both the off-diagonal operator entry $\tilde{K}_1$ and the free energy $\tilde{f}$ are very different from the long-range model, as could be anticipated. Particularly, while the long-range model has a double-well free energy, which suggests that the system may  exhibit phase separation, the predicted free energy is clearly single-welled and is not suggestive of phase separation. This observation is in agreement with the macroscopic evolution shown in Figs.~\ref{fig:JL2.2_L2}d-\ref{fig:JL2.2_L2}f. Here the initial profile is chosen as a triangular wave with average density $\rho_{ave} = 0.5$, amplitude $A=0.4$ and frequency $f=3$ and the simulation is run for a total time of $t_{tot}=0.2$ and a timestep of $\Delta t_{sim} = \Delta x_\gamma^2 / 2 = 2 \times 10^{-4}$. While the long-range model reached a steady-state with a bistable profile, the Stat-PINNs prediction decays to the average value, in agreement with the particle simulations. Although the results from PINNs capture the overall trend of the macroscopic evolution, noticeable errors exist at the beginning of the simulations.  

We remark that in the three particle processes discussed, the errors observed in the macroscopic dynamics predicted by PINNs can be eliminated by increasing the number of time intervals $\Delta t$ used for training the networks (see Appendix~\ref{Append:LargeData}), although this will naturally entail an increased computational cost. This therefore indicates that statistical mechanics relations can increase the robustness of Stat-PINNs, particularly when the amount of data is limited and the noise is not negligible. In any case, macroscopic data, regardless of how extensive this is, is insufficient to characterize the physical free energy and dissipative operator in the examples considered.


Finally, we would like to comment on the computational cost  for Stat-PINNs and the KMC particle simulations, highlighted in Table~\ref{Table:Computation_Cost}. As it is to be expected, direct particle simulations of long-time macroscopic evolutions are extremely costly, even for the one-dimensional example chosen. Here, parallelization is only done for independent realizations (600 in this case), as causality severely hinders parallelizations in time. For the specific particle systems chosen, the long-time KMC simulations performed took of the order of 1 to 6 days at full parallelization (or 1 to 10 years without any parallelization).  
With the Stat-PINNs strategy, the training and macroscopic prediction can all be done within 30 minutes. Here, the computational cost is dominated by the data collection process, which consists of a large number (28 profiles times $R=10^4$ realizations) of independent short-time trajectories. For the most ideal case, where independent trajectories are fully parallelized,  data collection can be done within a minute. And even in the most undesired case, where no parallelization is available, the computational cost of Stat-PINNs is still at least one to two orders of magnitude faster than particle simulations. Furthermore, Stat-PINNs can be applied to predict the evolution for different initial profiles and different boundary conditions, at minimal cost. Finally we want to comment that although all particle simulations mentioned above are  performed using the BKL algorithm, other algorithms with more efficient searching \citep{Schulze2008Efficient} or parallelization strategies \citep{Arampatzis2011Hierarchical} could be applied to further reduce the computational cost for the data collection in Stat-PINNs.

\section{Conclusions and Discussions}
\label{Sec:Conclusion}

In this work, we proposed a machine learning architecture called Statistical-Physics-Informed Neural Networks (Stat-PINNs) for learning purely dissipative evolution equations of GENERIC type from short-time particle simulations. It consists of two neural networks that are sequentially trained to learn the unique discretized dissipative operator and free energy density of the system, through a carefully designed structure-preserving parameterization method. More specifically, the proposed strategy strongly ensures that the dissipative operator is symmetric (as required by Onsager's reciprocity relations), positive semi-definite (in accordance with the second law of thermodynamics), and that the ensuing dynamics are conserved, when applicable, as is the the case of diffusive phenomena. The key idea behind the proposed framework is that field fluctuations, in addition to the average particle dynamics, contain crucial information for tackling the inverse problem of thermodynamic model discovery from data. Such fluctuations are leveraged in two ways. First, the fluctuation-dissipation relation is used to uniquely determine the dissipative operator. Second, the evolution equation is treated as a stochastic differential equation when defining the loss function used for learning the free energy. That is, in contrast to traditional PINNs, the loss function contains a denominator that is related to the stochastic noise estimated from statistical physics to normalize the equation's residue. The Stat-PINNs architecture is demonstrated over a particle process with three different Arrhenius-type interactions, of which only one has an analytically known macroscopic evolution equation. The predicted continuum evolution from Stat-PINNs perfectly agrees with the true particle dynamics in all cases, while PINNs exhibits noticeable deviations from the true evolution. Although such deviations can be reduced by increasing the length of the particle simulations, and gathering more training data along the particle trajectories (see Appendix~\ref{Append:LargeData}), the unique free energy and operator of the system cannot be learned by PINNs. Stat-PINNs may thus be regarded as more physically interpretable and also more robust when temporal data from particle simulations is limited.

\section*{Acknowledgment}

S.~H., Z.~H.~and C.~R.~gratefully acknowledge support from NSF CMMI-2047506, (MRSEC) DMR-1720530, and the US Department of the Army W911NF2310230. J.~Z.~also acknowledges support from W911NF2310230. The authors further thank Mr.~Travis Leadbetter for his valuable insight to encode the positive semi-definiteness in the proposed architecture, and Prof.~Markos Katsoulakis for suggesting the Arrhenius process as an important case study and sharing his knowledge on the coarse-graining of this particle system. 

\appendix

\section{Discussion on the Nonlocal Operator Beyond the Tri-diagonal Form}
\label{Append:NonlocalK}

When the dissipative operator $\K_z$ is nonlocal in the thermodynamic force $Q$, the discretized operator is no longer a tri-diagonal matrix and more non-zero off-diagonal entries need to be considered. In this case, the parameterization method introduced in  Sec.~\ref{Sec:K_Discrete} may still be utilized. In general, we can assume that the operator entry $\left\langle \gamma_{j}, \K_{z} \gamma_{i} \right\rangle \neq 0$ only when $j \in \{i, i \pm 1, \cdots, i \pm n_K \}$ with $n_K$ being a positive integer. Then, all non-zeros operator entries can be written as
\begin{equation}
\label{Eq:Kp_Zp}
    K_p(\mathbf{Z}_i^p) 
    = \left\langle \gamma_{i+p}, \K_z \gamma_{i} \right\rangle 
    = \left\langle \gamma_{i}, \K_z \gamma_{i+p} \right\rangle,
\end{equation}
where $p=0,1,\cdots,n_K$ and $\mathbf{Z}_{i}^p$ represents the profile $z$ at multiple local points within the non-local dependence on $z$ for calculating the integral.
Furthermore, when $z$ satisfies a conservation law, $K_0$ can be expressed as
\begin{equation}
\label{Eq:K0_MassCons_123}
     K_0(\mathbf{Z}^0_i)
     = - \sum_{p=1}^{n_K} \left[ K_p(\mathbf{Z}^p_{i-p}) 
     + K_p(\mathbf{Z}^p_i) \right],
\end{equation}
analogously to Eq.~\ref{Eq:K0_MassCons} for the fully local case. Then, $n_K$ neural networks, representing the $n_K$ off-diagonal entries would be required to learn the operator.

As mentioned in Sec.~\ref{Sec:K_Discrete}, when $z$ obeys a conservation law and $\K_z$ is local on $Q$ (indicating a tri-diagonal form), the weak form of the evolution equation can be written as Eq.~\ref{Eq:Weak_T-GENERIC_MassCons}, indicating a diffusive equation. 
In comparison, when $\K_z$ is nonlocal on $Q$, the evolution equation (Eq.~\ref{Eq:Weak_T-GENERIC}) can be re-written as
\begin{equation}
\begin{split}
\label{Eq:Weak_T-GENERIC_MassCons_Nonlocal-Q}
    \sum_{a} & \left\langle \gamma_{b}, \gamma_{a} \right\rangle
    \frac{\partial z_a}{\partial t} 
    \\
    & = - \sum_{p = 1}^{n_K} 
    \left\langle \gamma_{b+p}, \K_{z} \gamma_{b} \right\rangle 
    \left( Q_{b+p} - Q_b \right)
    \\
    & \quad 
    + \sum_{p = 1}^{n_K} 
    \left\langle \gamma_{b}, \K_{z} \gamma_{b-p} \right\rangle 
    \left( Q_{b} - Q_{b-p} \right) \\
    & = -\sum_{p = 1}^{n_K} 
    \Delta_b^{(p)} \left[ \left\langle \gamma_{b+p}, \K_{z} \gamma_{b} \right\rangle 
    \Delta_f^{(p)} Q_b \right],
\end{split}
\end{equation}
where $\Delta_f^{(p)} B_i = B_{i+p} - B_i$ is the $p$-step forward finite difference for an arbitrary variable $B$ and $\Delta_b^{(p)} B_i = B_{i} - B_{i-p}$ is the $p$-step backward finite difference. Equation~\ref{Eq:Weak_T-GENERIC_MassCons_Nonlocal-Q} can be viewed as a generalized version of Eq.~\ref{Eq:Weak_T-GENERIC_MassCons} with nonlocal diffusion behavior, where different off-diagonal entries serve as mobility functions corresponding to different nonlocal spatial derivatives.
Equation~\ref{Eq:Weak_T-GENERIC_MassCons_Nonlocal-Q} can also be written in matrix form as
\begin{equation}
\label{Eq:Discrete_T-GENERIC_MassCons_Matrix}
   \mathbf{M \dot{z}}
    = - \sum_{p = 1}^{n_K}  \left[
    \mathbf{D}^{(p)} \mathbf{\Lambda}^{(p)} 
    \left( - \mathbf{D}^{(p) \mathbf{T}} \right) \right] \mathbf{Q},
\end{equation}
where
\begin{equation*}
    \mathbf{z} = 
    \begin{pmatrix}
        z_1 \\
        z_2 \\
        \vdots \\
        z_{N_\gamma}
    \end{pmatrix},
    \quad
    \mathbf{Q} = 
    \begin{pmatrix}
        Q_1 \\
        Q_2 \\
        \vdots \\
        Q_{N_\gamma}
    \end{pmatrix},
\end{equation*}
\begin{equation*}
    \mathbf{M} = 
    \begin{pmatrix}
        \langle \gamma_1, \gamma_1 \rangle
        & \langle \gamma_1, \gamma_2 \rangle & \cdots 
        & \langle \gamma_1, \gamma_{N_\gamma} \rangle  \\
        \langle \gamma_2, \gamma_1 \rangle
        & \langle \gamma_2, \gamma_2 \rangle & \cdots 
        & \langle \gamma_2, \gamma_{N_\gamma} \rangle \\
        \vdots  & \vdots & \ddots & \vdots \\
        \langle \gamma_{N_\gamma}, \gamma_1 \rangle
        & \langle \gamma_{N_\gamma}, \gamma_2 \rangle & \cdots 
        & \langle \gamma_{N_\gamma}, \gamma_{N_\gamma} \rangle
    \end{pmatrix},
\end{equation*}
\begin{equation*}
\begin{split}
    \mathbf{\Lambda}^{(p)} = 
    \begin{pmatrix}
        K_{1+p,1}
        & 0 & \cdots & 0  \\
        0 & K_{2+p,2}& \ddots & \vdots \\
        \vdots  & \ddots & \ddots & 0 \\
        0 & \cdots & 0 & K_{p N_{\gamma}}
    \end{pmatrix}
\end{split}
\end{equation*}
with $K_{ji} = \left\langle \gamma_{j}, \K_{z} \gamma_{i} \right\rangle$, and $\mathbf{D}^{(p)} = \left( D^{(p)}_{ij} \right)_{N_{\gamma} \times N_{\gamma}}$, with
\begin{equation*}
    D^{(p)}_{ij} 
    = \left\{ \begin{aligned}
    & 1, 
    \quad
    && i = j\\
    & -1,
    && i = j+p \text{ or } i = j+p - N_{\gamma}\\
    & 0
    && \text{otherwise} \\
    \end{aligned} \right. 
\end{equation*}
Here $\mathbf{D}^{(p)}$ and $-\mathbf{D}^{(p) \mathbf{T}}$ represent the  $p$-step backward and forward finite difference matrices, respectively. We note that we have applied periodic boundary conditions here for simplicity.

Finally, the discretized operator should be positive semi-definite. 
If the operator is local on $Q$, (being a tridiagonal form), a positive semi-definite $\K_z$ requires
\begin{equation}
\begin{split}
\label{Eq:PSD_tridiagonal}
    & \mathbf{u^T D}^{(1)} \mathbf{\Lambda}^{(1)} \left(- \mathbf{D}^{(1)\mathbf{T}} \right) \mathbf{u} 
    \\
    & \quad 
    = -\left(  \mathbf{D}^{(1)\mathbf{T}}  \mathbf{u} \right)^{\mathbf{T}} \mathbf{\Lambda}^{(1)}  \left( \mathbf{D}^{(1)\mathbf{T}} \mathbf{u} \right) \geq 0,
\end{split}
\end{equation}
for an arbitrary vector $\mathbf{u}$. 
This can be simplified to $K_1(\mathbf{Z}^1_{i}) \leq 0$ when $\K_z$ is local on the thermodynamic force $Q$. Such a practical strategy to encode the positive semi-definiteness when $\K_z$ is nonlocal on $Q$ has not yet been discovered.

\section{Multi-point Expansion and Middle-point Expansion for the Discretized Operator}
\label{Append:MP&MidP}

In Sec.~\ref{Sec:K_Discrete}, we introduced the multi-point expansion method to parameterize the discrete operator entry as $K_p(\mathbf{Z}_i^p) = \left \langle \gamma_{i+p}, \K_z \gamma_i \right \rangle = \left \langle \gamma_i, \K_z \gamma_{i+p} \right \rangle$. 
This method is also mathematically equivalent to expanding the same entry at the middle point $x_{i+\frac{p}{2}} = (x_i+x_{i+p})/2$,
\begin{equation}
\label{Eq:Kp_MP_to_Mid}
    K_p(\mathbf{Z}^p_i) 
    = \tilde{K}_p(\tilde{\mathbf{Z}}^p_{i+\frac{p}{2}}), 
\end{equation}
where $\tilde{K}_p$ represents the middle-point expansion form of $K_p$, and $\tilde{\mathbf{Z}}^p_{i+\frac{p}{2}}$ represents the corresponding values of the local profile and its derivatives.
For example, when $\K_z$ is local on both $z$ and $Q$, the two operator entries $K_0$ and $K_1$ can be equivalently written as
\begin{equation}
\begin{gathered}
\label{Eq:K0_K1_MP_to_Mid}
    K_0(z_{i-1}, z_i, z_{i+1}) = \tilde{K}_0(z_i, \left. \nabla z \right|_{i}, \left. \nabla^2 z \right|_{i}), \\
    K_1(z_{i}, z_{i+1}) = \tilde{K}_1(z_{i+\frac{1}{2}}, \left. \nabla z \right|_{i+\frac{1}{2}}), 
\end{gathered}
\end{equation}
with
\begin{equation*}
\begin{gathered}
    \left. \nabla z \right|_{i} = \frac{z_{i+1} - z_{i-1}}{2 \Delta x_\gamma},
    \quad \quad
    \left. \nabla^2 z \right|_{i} 
    = \frac{ z_{i+1} - 2 z_i + z_{i-1} }{\Delta x_\gamma^2} , \\
    \text{and} \quad
    z_{i+\frac{1}{2}} = 
    \frac{z_{i} + z_{i+1} }{2},
    \quad \quad 
    \left. \nabla z \right|_{i+\frac{1}{2}} 
    = \frac{z_{i+1} - z_{i} }{\Delta x_\gamma} .
\end{gathered}
\end{equation*}

\section{Numerical Schemes for the Free Energy and Thermodynamic Force}
\label{Append:fe_Q}

In Sec.~\ref{Sec:fe_Discrete}, we introduced the expression of the thermodynamic force $Q$ when the free energy has different functional dependencies. In this section, we will discuss the expression of $f$ and $Q$ in the discrete setting. 

\begin{figure*}
\centering
\includegraphics[width=0.85\linewidth]{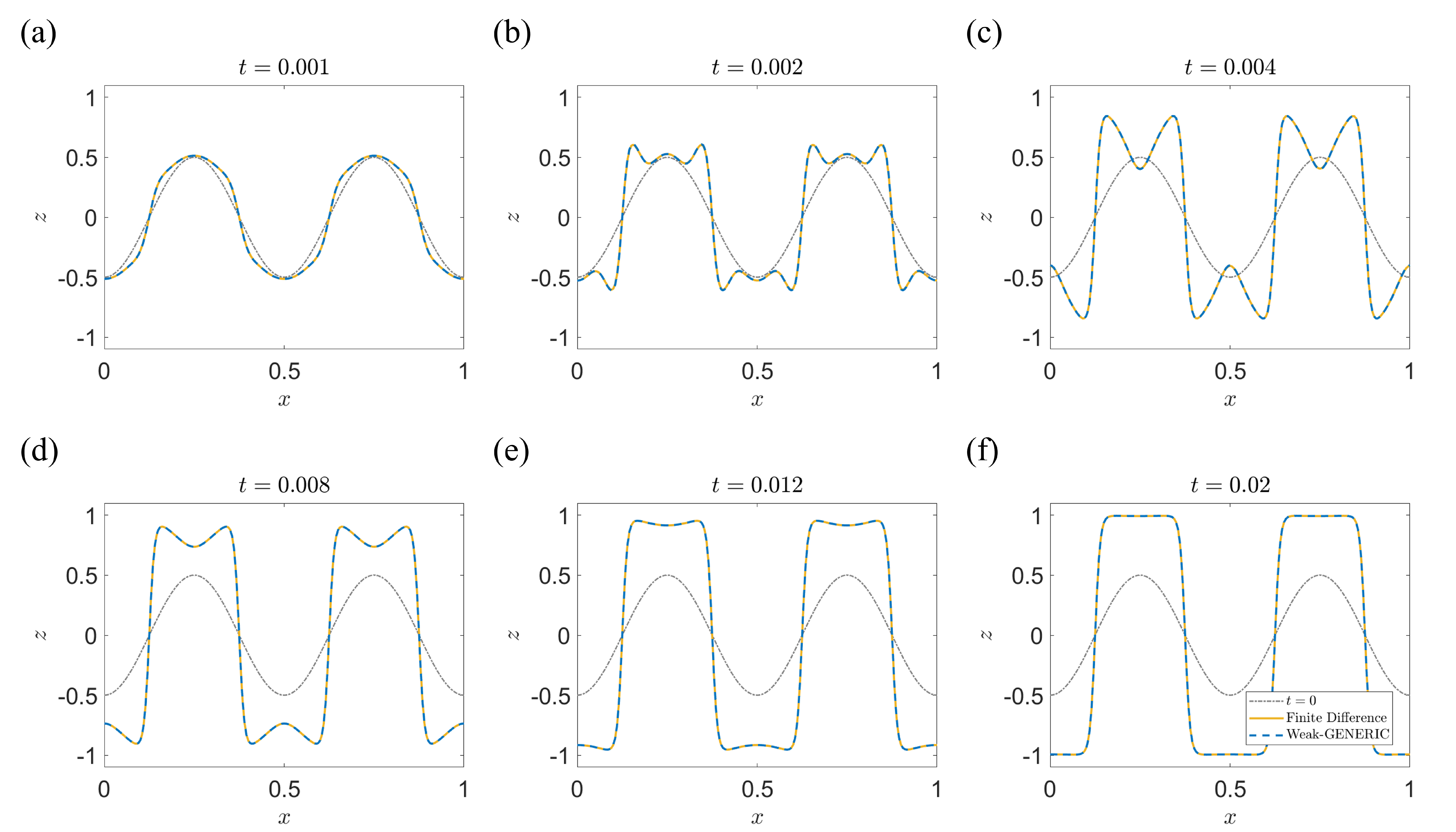}
\caption{Solution of a 4th-order Cahn-Hilliard equation using a finite difference scheme (yellow solid) and the proposed weak-GENERIC method with linear shape functions (blue dashed). (a-f) Different snapshots for the evolution of $z(x,t)$ at $t=0.001, 0.002, 0.004, 0.008, 0.012$ and 0.02. The black dash dotted lines indicate the sinusoidal initial profile at $t=0$.}
\label{fig:4th-CH}
\end{figure*}

Here we assume that the free energy density can be written as $f = f(z, \nabla z, \nabla^2 z)$. If we use a backward scheme for the gradient, $\left. \nabla z\right|_i = (z_i - z_{i-1}) / \Delta x_\gamma$, and a central scheme for the Laplacian, $\left. \nabla^2 z\right|_i = (z_{i-1} - 2 z_i + z_{i+1}) / \Delta x_\gamma ^2$, and denote $f_i = f(z_i, \left. \nabla z \right|_i, \left. \nabla^2 z \right|_i)$, the variation of the total free energy functional can be given by
\begin{equation}
\begin{split}
\label{Eq:delta_F}
    \delta F 
    & = \sum_{i=1}^{N_\gamma} \left( 
    f_{i,z} \delta z_i
    + f_{i, \nabla z} 
    \frac{ \delta z_i - \delta z_{i-1} }{\Delta x_\gamma}
    \right.
    \\
    & \quad
    \left.
    + f_{i, \nabla^2 z} 
    \frac{ \delta z_{i+1} - 2 \delta z_i + \delta z_{i-1} }{\Delta x_\gamma^2}
    \right) \Delta x_\gamma \\
    & = \sum_{i=1}^{N_\gamma} \left[ 
    f_{i,z} \delta z_i
    + \left( f_{i, \nabla z} 
    - \frac{f_{i, \nabla^2 z}}{\Delta x_\gamma} \right)
    \frac{ \delta z_i - \delta z_{i-1} }{\Delta x_\gamma}
    \right.
    \\
    & \quad
    \left.
    + \frac{f_{i, \nabla^2 z}}{\Delta x_\gamma} 
    \frac{ \delta z_{i+1} - \delta z_i}{\Delta x_\gamma}
    \right] \Delta x_\gamma,
\end{split}
\end{equation}
where $f_{i,z} = \partial f_i / \partial z$ and similarly for $f_{i,\nabla z}$ and $f_{i,\nabla^2 z}$. For simplicity, we denote $r_i = f_{i, \nabla z} - f_{i, \nabla^2 z} / \Delta x_\gamma$. Then, the summation of the second term in the square bracket above can be given by
\begin{equation}
\begin{split}
    & \sum_{i=1}^{N_\gamma}
    \left( r_i \frac{ \delta z_i - \delta z_{i-1} }{\Delta x_\gamma} \right) \Delta x_\gamma
    \\
    & = 
    \sum_{i=1}^{N_\gamma} \left( 
    \frac{ r_i \delta z_i - r_{i-1} \delta z_{i-1} }{\Delta x_\gamma}
    + \frac{ r_{i-1} \delta z_{i-1} - r_{i} \delta z_{i-1} }{\Delta x_\gamma}
    \right) \Delta x_\gamma \\
    & = 
    \frac{ r_{N_\gamma} \delta z_{N_\gamma} - r_{0} \delta z_{0} }{\Delta x_\gamma}
    + \sum_{i=1}^{N_\gamma} \left( 
    \frac{ r_{i-1} \delta z_{i-1} - r_{i} \delta z_{i-1} }{\Delta x_\gamma}
    \right) \Delta x_\gamma \\
    & =  
    - \sum_{i=1}^{N_\gamma} \left( 
    \frac{ r_{i+1} - r_{i} }{\Delta x_\gamma}
    \right) \delta z_{i}  \Delta x_\gamma, \\
\end{split}
\end{equation}
where the first term in the second line is cancelled out when applying  periodic boundary conditions for $z$. Similarly, the third term in the square bracket of Eq.~\ref{Eq:delta_F} can be expressed as
\begin{equation}
\begin{split}
    & \sum_{i=1}^{N_\gamma}
    \left( \frac{f_{i, \nabla^2 z}}{\Delta x_\gamma}  \frac{ \delta z_{i+1} - \delta z_{i} }{\Delta x_\gamma} \right) \Delta x_\gamma
    \\
    & \quad 
    =  
    - \sum_{i=1}^{N_\gamma} \left( 
    \frac{ f_{i, \nabla^2 z} - f_{i-1, \nabla^2 z} }{\Delta x_\gamma^2}
    \right) \delta z_{i}  \Delta x_\gamma.
\end{split}
\end{equation}
Combining the above results, Eq.~\ref{Eq:delta_F} can be simplified as
\begin{equation}
\begin{split}
    \delta F
    & =   
    \sum_{i=1}^{N_\gamma} \left( 
    f_{i,z}
    - \frac{ r_{i+1} - r_{i} }{\Delta x_\gamma}
    - \frac{ f_{i, \nabla^2 z} - f_{i-1, \nabla^2 z} }{\Delta x_\gamma^2}
    \right) \delta z_{i}  \Delta x_\gamma \\
    & =   
    \sum_{i=1}^{N_\gamma} \left( 
    f_{i,z}
    - \frac{ f_{i+1, \nabla z} - f_{i, \nabla z} }{\Delta x_\gamma}
    + \frac{ f_{i+1, \nabla^2 z} - f_{i, \nabla^2 z} }{\Delta x_\gamma^2}
    \right.
    \\
    & \quad
    \left.
    - \frac{ f_{i, \nabla^2 z} - f_{i-1, \nabla^2 z} }{\Delta x_\gamma^2}
    \right) \delta z_{i}  \Delta x_\gamma \\
    & =   \sum_{i=1}^{N_\gamma} \left( 
    f_{i,z}
    - \frac{ f_{i+1, \nabla z} - f_{i, \nabla z} }{\Delta x_\gamma}
    \right.
    \\
    & \quad
    \left.
    + \frac{ f_{i+1, \nabla^2 z} 
    - 2 f_{i, \nabla^2 z} 
    + f_{i-1, \nabla^2 z}}{\Delta x_\gamma^2}
    \right) \delta z_{i}  \Delta x_\gamma.
\end{split}
\end{equation}
Therefore, the discrete thermodynamic force at $x_i$ is given by
\begin{equation}
\begin{split}
    Q_i 
    & =  f_{i,z}
    - \frac{ f_{i+1, \nabla z} - f_{i, \nabla z} }{\Delta x_\gamma}
    \\
    & \quad 
    + \frac{ f_{i+1, \nabla^2 z} 
    - 2 f_{i, \nabla^2 z} 
    + f_{i-1, \nabla^2 z}}{\Delta x_\gamma^2}
    \\
    & \quad
    = f_{i,z}
    - \frac{\Delta_f f_{i, \nabla z}}{\Delta x_\gamma}
    + \frac{\Delta_f \Delta_b f_{i, \nabla^2 z}}{\Delta x_\gamma^2}.
\end{split}
\end{equation}
Here, we want to emphasize that assuming that the free energy has a functional dependence on a backward scheme gradient will lead to a forward scheme spatial derivative of $f_{i,\nabla z}$ in the expression of $Q$. Similarly, a forward scheme for $\nabla z$ will lead to a backward scheme for $\nabla f_{,\nabla z}$, and a central scheme for $\nabla z$ will lead to a central scheme for $\nabla f_{,\nabla z}$.

\section{Solving Higher-Order PDEs with Linear Shape Functions}
\label{App:Cahn-Hilliard}

In Sec.~\ref{Sec:fe_Discrete}, we have theoretically shown the possibility of solving PDEs with order higher than two only using linear shape functions (as long as the operator is second order). In this appendix, we will numerically verify such claim.

Consider a one-dimensional diffusion problem for field $z(x,t)$ described by a Laplacian dissipative operator $\K=-\nabla^2$ and the following free energy density,
\begin{equation}
    f(z, \nabla z, \nabla^2 z) 
    = \frac{1}{4} \left( z^2 - 1 \right)^2
    + \frac{\alpha_1}{2} \left(\nabla z \right)^2
    + \frac{\alpha_2}{2} \left(\nabla^2 z \right)^2,
\end{equation}
which includes a quartic double-well term, a quadratic gradient term, and a quadratic term for the Laplacian of $z$, where $\alpha_1$ and $\alpha_2$ are two coefficients. The evolution equation, Eq.~\ref{Eq:T-GENERIC}, for $z$ can be written as the following 6th-order equation
\begin{equation}
\label{Eq:6thCH-GL}
    \frac{\partial z}{\partial t} 
    = \nabla^2 \left( z^3 - z
    - \alpha_1 \nabla^2 z 
    + \alpha_2 \nabla^4 z \right).
\end{equation}

Here we solve Eq.~\ref{Eq:6thCH-GL} by two methods. The first is the standard finite difference method, which we will here take as the ground truth solution. The second corresponds to the weak form of the GENERIC formalism using linear shape functions as in Eq.~\ref{Eq:Weak_T-GENERIC}, which we denote as weak-GENERIC. In both cases, a simple forward Euler scheme is used for the time derivative and a 2nd-order central scheme is used for the Laplacian. The 4th-order gradient $\nabla^4$ is obtained by applying the Laplacian twice.


\begin{figure*}
\centering
\includegraphics[width=0.85\linewidth]{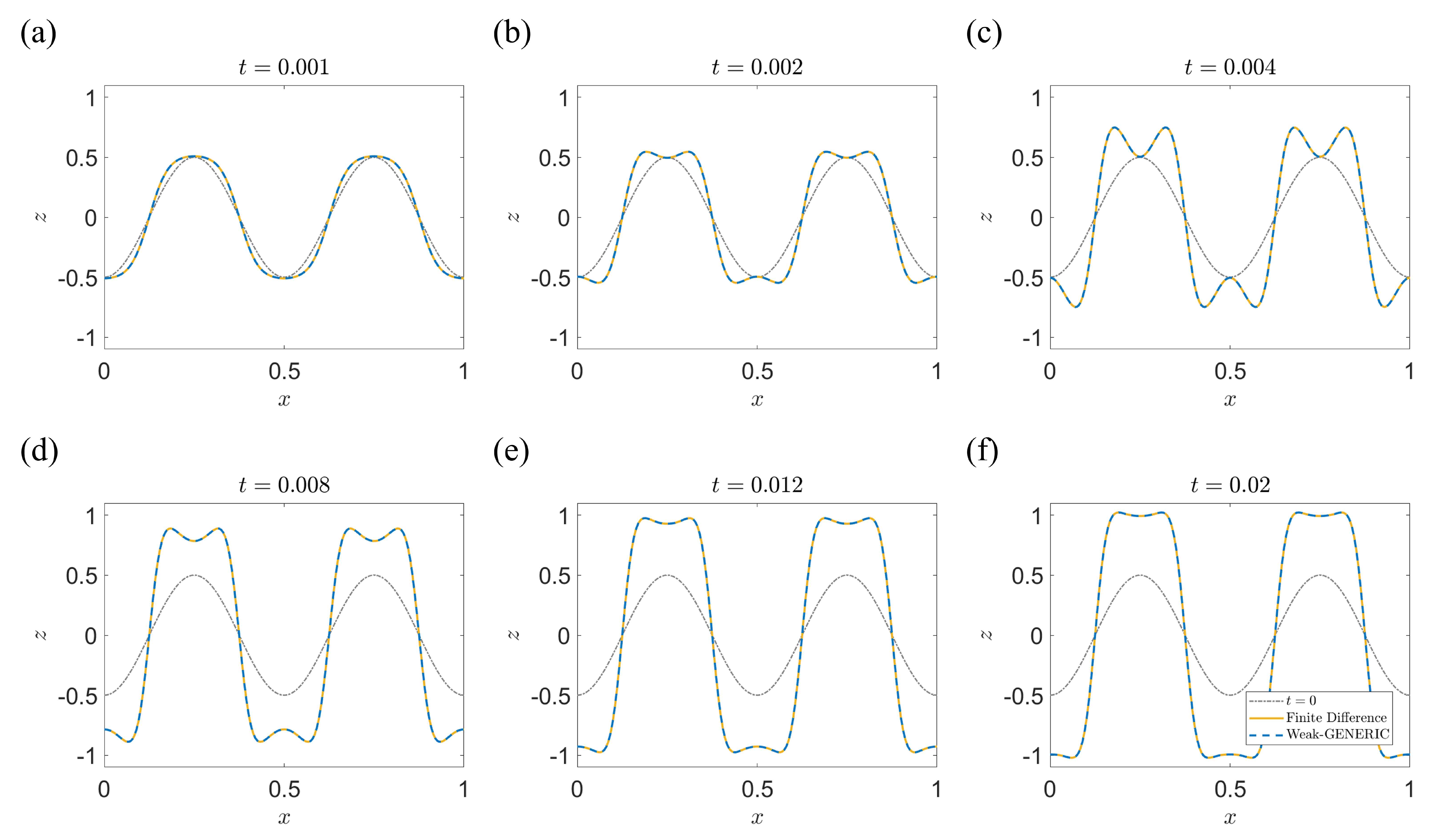}
\caption{Solution of a 6th-order evolution equation using a finite difference scheme (yellow solid) and the proposed weak-GENERIC method with linear shape functions (blue dashed). (a-f) Different snapshots for the evolution of $z(x,t)$ at $t=0.001, 0.002, 0.004, 0.008, 0.012$ and 0.02. The black dash dotted lines indicate the sinusoidal initial profile at $t=0$.}
\label{fig:6th-CH}
\end{figure*}

Figures~\ref{fig:4th-CH} and \ref{fig:6th-CH} depict the numerical results from two examples using different coefficients $\alpha_1$ and $\alpha_2$. In Fig.~\ref{fig:4th-CH}, we choose $\alpha_1 = 10^{-4}$ and $\alpha_2 = 0$, indicating a 4th-order Cahn-Hilliard equation. The equation is discretized using 200 spatial nodes and solved with a timestep $\Delta t = 2.5 \times 10^{-8}$. In Fig.~\ref{fig:6th-CH}, we choose $\alpha_1 = 10^{-4}$ and $\alpha_2 = 10^{-5}$, indicating a 6th-order diffusion equation. The equation is discretized using 200 spatial nodes and solved with a timestep $\Delta t = 10^{-9}$. In both examples, the equations are solved from the same sinusoidal initial condition $z = 0.5 \sin (2 \pi x)$, plotted as a black dash dotted line. In both examples, the solution obtained using the weak-GENERIC method (blue dashed lines) agree very well with that obtained using a finite difference scheme (yellow solid lines). This indicates that the proposed approach utilizing linear shape functions can be used to solve PDEs with order higher than two, as long as the dissipative operator is at most of second order.

\section{Structure Details for Stat-PINNs}
\label{Append:Stat-PINNs_Details}

When learning the off-diagonal entry $K_1$ from neural network $\text{NN}_{K_1}$, a normalization of the input data is needed. First, the local field values $\mathbf{Z}^1$ at multiple points are transformed into $\tilde{\mathbf{Z}}^1$, which consists of local field values and its derivatives at the middle point as discussed in Appendix~\ref{Append:MP&MidP}. Then, each component of $\tilde{\mathbf{Z}}^1$ is subtracted by the mean and then divided by the standard deviation of that component within all training data to obtain the normalized input $\tilde{\mathbf{Z}}^{1*}$ for $\text{NN}_{K_1}$. 
This normalization process can decouple the dependence that naturally exists between sequential points (e.g., between $z_i$ and $z_{i+1}$), increasing the training robustness, and making it easier to understand and manipulate practically. 
The off-diagonal entry $K_1$ can then by parameterized by
\begin{equation}
\label{Eq:K1_NNK1}
    K_{1} \left( \mathbf{Z}^1; \pmb{\theta}_1 \right)
    = \left| \mu_{K_1} \right| g\left( 
    \text{NN}_{K_1} \left( 
    \tilde{\mathbf{Z}}^{1*}; \pmb{\theta}_1 \right) \right),
\end{equation}
where $\mu_{K_1}$ is the mean of all measured off-diagonal entry data for ensuring that the output of $\text{NN}_{K_1}$ is of scale one, and $g(\cdot)$ is a non-positive function to encode the positive semi-definiteness of the operator $\K_z$. Here, we choose $g(y) = - y - e^{-5}$ if $y \geq 0$ and $g(y) = - e^{y-5}$ if $y<0$ \citep{sivaprasad2021curious, huang2022variational}. 

A similar normalization process is applied to the second neural network $\text{NN}_f$. Each component of $\tilde{\mathbf{Z}}^f$, the local profile and its derivatives for expressing the free energy, is subtracted by the mean and then divided by the standard deviation of that component within all training data to obtain the normalized input $\tilde{\mathbf{Z}}^{f*}$. The free energy density can then by represented by
\begin{equation}
\label{Eq:fe_NNf}
    f \left( \tilde{\mathbf{Z}}^f; \pmb{\theta}_f \right)
    = f^{\dag} 
    \text{NN}_{f} \left( 
    \tilde{\mathbf{Z}}^{f*}; \pmb{\theta}_f \right),
\end{equation}
where $f^{\dag}$ is an estimated scale for the free energy density based on the evolution equation in order to control that the output of $\text{NN}_f$ is of scale one. For the case of $f=f(z)$, we use $f^{\dag} = \mu_{\mathbf{M}} \mu_{\left| \dot{z} \right|} \mu_{\left|z\right|} / \mu_{K_1}$, where $\mu_{\mathbf{M}} = \left( \langle \gamma_{i-1}, \gamma_i \rangle + \langle \gamma_{i}, \gamma_i \rangle  + \langle \gamma_{i+1}, \gamma_i \rangle \right) / 3$, $\mu_{\left| \dot{z} \right|}$ is the mean of the absolute value of the profile time rate, $\left| \Delta z/\Delta t \right|$, within all the training data, and $\mu_{\left|z\right|}$ is the mean of the absolute value of all the profile data. For the case of $f=f(z, \nabla z)$, we use $f^{\dag} = \mu_{\mathbf{M}} \mu_{\left| \dot{z} \right|} \left( \mu_{\left|z\right|} + \mu_{\left|\nabla z\right|} \Delta x_\gamma \right) / \mu_{K_1}$, where $\mu_{\left|\nabla z\right|}$ is the mean of the absolute value of all $\nabla z$.  For the case of $f=f(z, \nabla^2 z)$, we use $f^{\dag} = \mu_{\mathbf{M}} \mu_{\left| \dot{z} \right|} \left( \mu_{\left|z\right|} + \mu_{\left|\nabla^2 z\right|} \Delta x_\gamma^2 \right) / \mu_{K_1}$, where $\mu_{\left|\nabla^2 z\right|}$ is the mean of the absolute value of all $\nabla^2 z$. Finally, we want to emphasize that the training data for $\text{NN}_f$ are $\mathbf{Z}^Q$ and $\Delta \mathbf{Z}^Q$ with superscript $Q$. This is due to the fact that $Q$ is directly involved in the loss function Eq.~\ref{Eq:Loss_feNN} instead of $f$ and calculating $Q$ may require information from more local points. Therefore, a transformation from $\mathbf{Z}^Q$ to $\tilde{\mathbf{Z}}^f$ is also required at the beginning of training procedure.

\section{Variance Estimation for the Weak Form of the Evolution Equation}
\label{Append:Var_Weak_Eq}

As mentioned in Sec.~\ref{Sec:weak-GENERIC}, the added stochastic noise term for Eq.~\ref{Eq:Weak_T-GENERIC} is $\left\langle \gamma_{j}, \sqrt{ 2 \epsilon \mathcal{K}_{z}} \dot{W}_{x,t} \right\rangle$. We note that we are here omitting the subscript $\epsilon$ that emphasizes the dependence on the lattice size, for simplicity. 
Given that $\K_{z}$ is a linear, symmetric, positive semi-definite operator and $\dot{W}_{x,t}$ is a Gaussian process in space and time, the random variable $\left\langle \gamma_{j}, \sqrt{ 2 \epsilon \mathcal{K}_{z}} dW_{x,t} \right\rangle$ during a time interval $dt$ has a (multivariate) Gaussian distribution. The corresponding mean and variance, conditioned on the profile $z$ at time $t$, are given by
\begin{align}
\begin{split}
\label{Eq:Mean_Weak_T-GENERIC}
    & \mathbb{E} \left[ \left\langle \gamma_{j}, \sqrt{ 2 \epsilon \mathcal{K}_{z}} dW_{x,t} \right\rangle \middle| z \right]
    \\
    & \quad 
    =  \left\langle \gamma_{j}, \sqrt{ 2 \epsilon \mathcal{K}_{z}} 
    \mathbb{E}  \left[ dW_{x,t} \right]
    \right\rangle 
    = 0,
\end{split}
    \\
\begin{split}
\label{Eq:Var_Weak_T-GENERIC}
    & \text{Var} \left[ \left\langle \gamma_{j}, \sqrt{ 2 \epsilon \mathcal{K}_{z}} dW_{x,t} 
    \middle| z \right\rangle \right]
    \\
    & \quad 
    =  \mathbb{E} \left[ \left\langle \gamma_{j}, \sqrt{ 2 \epsilon \mathcal{K}_{z}} 
    dW_{x,t} 
    \right\rangle ^2 
    \middle| z \right] \\
    & \quad 
    =  2 \epsilon
    \int_x \int_y \sqrt{\mathcal{K}_{z}} \gamma_{j}(x) 
    \sqrt{\mathcal{K}_{z}} \gamma_{j}(y) 
    \mathbb{E} \left[ dW_{x,t} dW_{y,t} \right]
     dx\, dy \\
    & \quad 
    =  \int_x \int_y 2 \epsilon 
    \sqrt{\mathcal{K}_{z}} \gamma_{j}(x)  \sqrt{\mathcal{K}_{z}} \gamma_{j}(y) 
    \delta(x-y) dt\, dx\, dy \\
    & \quad 
    =  2 \epsilon \left\langle \sqrt{\mathcal{K}_{z}} \gamma_{j}, \sqrt{\mathcal{K}_{z}} 
    \gamma_b
    \right\rangle dt \\
    & \quad 
    = 2 \epsilon \left\langle \gamma_{j}, 
    \mathcal{K}_{z} \gamma_j
    \right\rangle dt,
\end{split}
\end{align}
where we have used 
$\mathbb{E} \left[ dW_{x,t} dW_{y,t} \right] = \delta(x-y) dt$. Therefore, considering the average evolution of $R$ realizations of particle dynamics during a short time interval $\Delta t$, the variance for the residue of Eq.~\ref{Eq:Weak_T-GENERIC} can be given by $ \sigma_{Eq,j}^2 = 2 \epsilon \langle \gamma_{j},  \mathcal{K}_{z} \gamma_j \rangle / \left(R \Delta t \right)$.

\section{Long-range Analytic Model for Arrhenius Process}
\label{Append:Arr_Ana}

The long-range analytic model is introduced in Sec.~\ref{Sec:Arr_Model} with Eqs.~\ref{Eq:ContEq_Arr_LongL}--\ref{Eq:ArrK}. Given that Eq.~\ref{Eq:ContEq_Arr_LongL} is a diffusion equation, the following numerical scheme is used for solving the continuum evolution
\begin{equation}
\label{Eq:Long-range_Scheme}
    \rho_i^{n+1} = \rho_i^n + \frac{m_{i+\frac{1}{2}}^n (Q_{i+1}^n - Q_i^n) - m_{i-\frac{1}{2}}^n (Q_{i}^n - Q_{i-1}^n) }{\Delta x_\gamma^2} \Delta t,
\end{equation}
where $\rho_i^n = \rho(x_i, t^n)$ with $t^n = n \Delta t$ and similarly for $Q$. We note that the mobility is evaluated at the mid-point to preserve the conservation of $\rho$ and that the density profiles at mid-points are defined as $\rho_{i+\frac{1}{2}}^n = (\rho_i^n + \rho_{i+1}^n) / 2$.

The discrete operator entries for the long-range Arrhenius analytic expression can be evaluated by the following calculation
\begin{equation}
\begin{split}
    \left\langle \gamma_{i}, \K_{\rho} \gamma_{i} \right\rangle
    & = 
    \left\langle \nabla \gamma_{i},  \cdot m\left[\rho\right] \nabla \gamma_{i} \right\rangle \\
    & = 
    \int_{x_{i-1}}^{x_{i+1}} 
    m[\rho(x)] \left( \nabla \gamma_i(x) \right)^2 dx \\
    & =
    \frac{1}{\Delta x_\gamma^2} 
    \int_{x_{i-1}}^{x_{i+1}} 
    \biggl[ m_i 
    + \left. \nabla m \right|_i \left( x - x_i  \right)
    \\
    & \quad
    \left.
    + \frac{1}{2} \left.\nabla^2 m \right|_i \left( x - x_i  \right)^2
    + O\left(\Delta x_\gamma^3\right)
    \right] dx \\
    & = 
    \frac{2 m_i}{\Delta x_\gamma}
    + \frac{1}{3} 
    \left.\nabla^{2} m \right|_i 
    \Delta x_\gamma 
    + O\left(\Delta x_\gamma^3\right),
\end{split}
\end{equation}
and
\begin{equation}
\begin{split}
    & \left\langle \gamma_{i+1}, \K_{\rho} \gamma_{i} \right\rangle
    \\
    & \quad 
    = 
    \left\langle \nabla \gamma_{i+1},  \cdot m\left[\rho\right] \nabla \gamma_{i} \right\rangle \\
    & \quad
    =
    - \frac{1}{\Delta x_\gamma^2} 
    \int_{x_{i}}^{x_{i+1}} 
    \biggl[ m_{i+\frac{1}{2}} 
    + \left. \nabla m \right|_{i+\frac{1}{2}} \left( x - x_{i+\frac{1}{2}}  \right)
    \\
    & \quad \quad
    \left.
    + \frac{1}{2} \left.\nabla^2 m \right|_{i+\frac{1}{2}} \left( x - x_{i+\frac{1}{2}}  \right)^2
    + O\left(\Delta x_\gamma^3\right)
    \right] dx \\
    & \quad 
    = 
    - \frac{m_{i+\frac{1}{2}}}{\Delta x_\gamma}
    - \frac{1}{6} 
    \left.\nabla^{2} m \right|_{i+\frac{1}{2}} 
    \Delta x_\gamma 
    + O\left(\Delta x_\gamma^3\right).
\end{split}
\end{equation}
We note that all diagonal entries are expanded at the middle point as described in Appendix~\ref{Append:MP&MidP}.
Furthermore, if we denote $V=J*\rho$, the nonlocal mobility functional can be simplified into a local mobility function, i.e., $m[\rho] = \tilde{m}(\rho, V) = D \rho (1-\rho) e^{-V}$. Given that $\tilde{m}(\rho, V)$ and its higher order derivatives are now local, the first two spatial gradients of $m$ can be given by
\begin{equation}
\label{Eq:dmdx_rho}
    \nabla m
    = \tilde{m}_{,\rho} \nabla \rho
    + \tilde{m}_{,V} \nabla V
    = \tilde{m}_{,\rho} \nabla \rho
    - \tilde{m} \nabla V,
\end{equation}
\begin{equation}
\label{Eq:d2mdx2_rho}
\begin{split}
    \nabla^2 m
    & = \tilde{m}_{,\rho\rho} \left( \nabla \rho \right)^2
    + 2 \tilde{m}_{,\rho V} \nabla \rho \nabla V
    + \tilde{m}_{,\rho} \nabla^2 \rho
    \\
    & \quad
    + \tilde{m}_{,VV} \left( \nabla V \right)^2
    + \tilde{m}_{,V} \nabla^2 V \\
    & = \tilde{m}_{,\rho\rho} \left( \nabla \rho \right)^2
    - 2 \tilde{m}_{,\rho} \nabla \rho \nabla V
    + \tilde{m}_{,\rho} \nabla^2 \rho
    \\
    & \quad
    + \tilde{m} \left[ \left( \nabla V \right)^2
    - \nabla^2 V \right],
\end{split}
\end{equation}
where $\tilde{m}_{,\rho} = D (1-2\rho) e^{-V}$ and $\tilde{m}_{,\rho\rho} = -2D e^{-V}$. 

In practice, we choose a step function interaction $J$ as described in Sec.~\ref{Sec:Results} and approximate   $V=J\ast \rho$ numerically based on the nonlocality of the system. 
In all three cases considered in Sec.~\ref{Sec:Results},
the nonlocal effect is insignificant, which is verified by the macroscopic evolution. Therefore we approximate the convolution by a zeroth-order expansion as $V=J\ast \rho \approx 2 J_0 L \rho$.

\section{Settings for the Particle Simulation}
\label{Append:KMC_Settings}

The particle simulations performed for the Arrhenius process are run using the Kinetic Monte-Carlo (KMC) method. Such simulations were used to (1) generate the training data for Stat-PINNs and (2) validate the long-time continuum evolution equations. All particle simulations are coded in C$++$.

During data generation for Stat-PINNs, 
the initial profiles should be carefully chosen according to two criteria. First, we would like to avoid an initial profile with a too wide density range. This is because the jumping probability in Arrhenius process has an exponential factor to local interaction. Then the corresponding time scale is approximately proportional to $e^{J_0 L \rho}$, which indicates that the time scale at different points in the same profile can be exponentially different which may lead to practical issues. Second, the generated data should be relatively uniform and dense in the input space for training purposes and avoid extrapolation during the prediction stage. Particularly, here we targeted the space of $\left( \rho_{i+\frac{1}{2}}, \left. \nabla \rho \right|_{i+\frac{1}{2}} \right)$ for learning the off-diagonal entry $\tilde{K}_1 \left( \rho_{i+\frac{1}{2}}, \left. \nabla \rho \right|_{i+\frac{1}{2}} \right) = K(\rho_i, \rho_{i+1})$ of a local operator. 

\begin{table*}
\centering
\begin{tabular}{|c | c | c | c|} 
 \hline
 Label & $f$ & $A$ & $\rho_{ave}$ \\ 
 \hline
 1-7 & 1 
 & $0.05 + 0.03 i$ with $i=1, \cdots, 7$ 
 & $0.5 - (-1)^i \times 0.6 \left( 0.5 - A \right)$ with $i=1, \cdots, 7$
 \\ 
 \hline
 8-14 & 1
 & 0.31 
 & $0.272 + 0.057$ with $i=1, \cdots, 7$ \\
 \hline
 15-21 & 2
 & $0.04 + 0.03 i$ with $i=1, \cdots, 7$ 
 & $0.5 + (-1)^i \times 0.84 \left( 0.5 - A \right)$ with $i=1, \cdots, 7$ \\
 \hline
 21-28 & 2
 & 0.29 
 & $0.272 + 0.057$ with $i=1, \cdots, 7$ \\
 \hline
\end{tabular}
\caption{Parameters used for the 28 initial cosine profiles, $\rho(x) = \rho_{ave} - A \cos (4 \pi f x)$, used in the KMC particle simulation.}
\label{Table:InitialProfile}
\end{table*}

\begin{figure}[h!]
\centering
\includegraphics[width=0.8\linewidth]{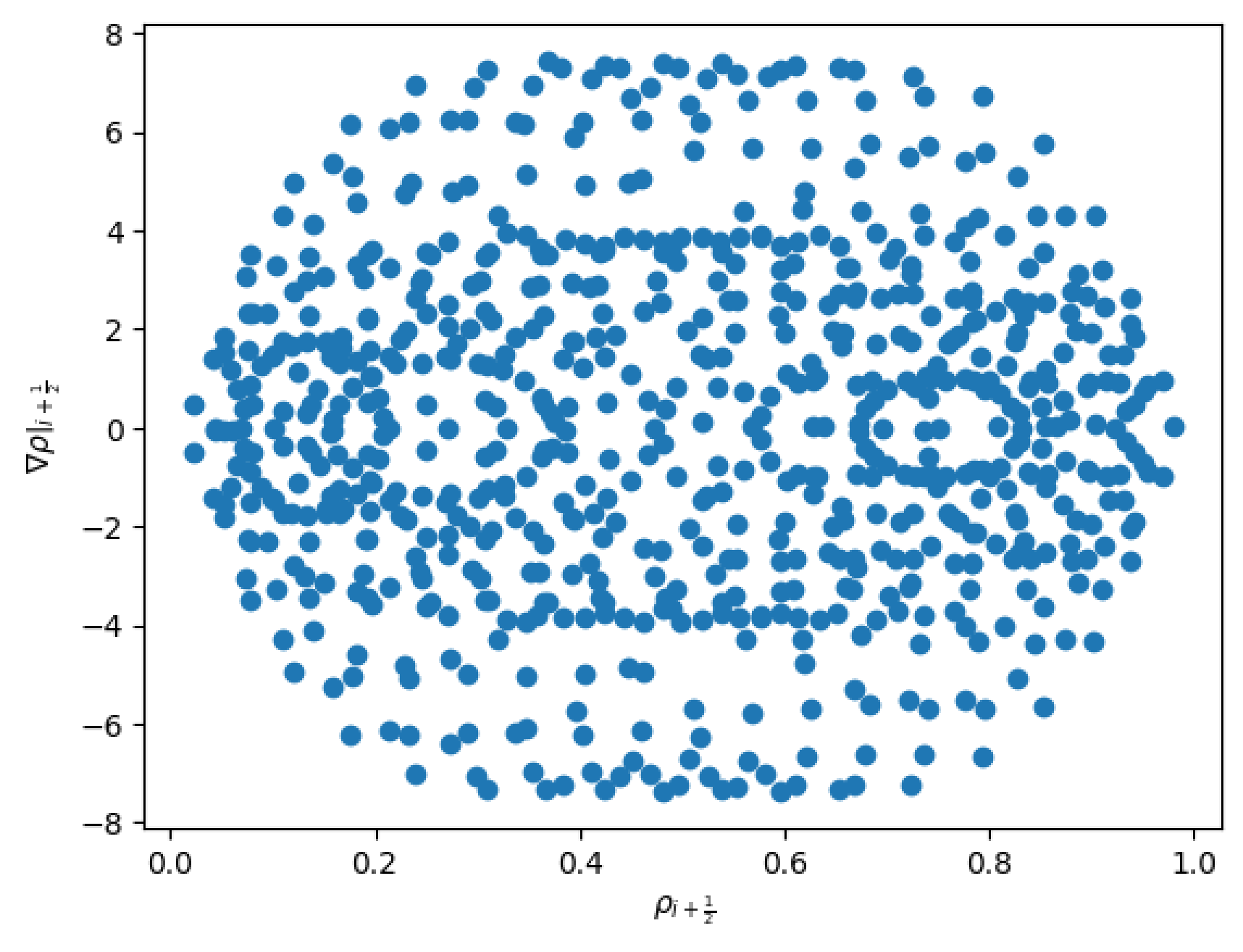}
\caption{Distribution of the training data in the space of $\left( \rho_{i+\frac{1}{2}}, \left. \nabla \rho \right|_{i+\frac{1}{2}} \right)$ generated from the 28 initial profiles.}
\label{fig:DataDistribution}
\end{figure}

\begin{figure}[h!]
\centering
\includegraphics[width=.9\linewidth]{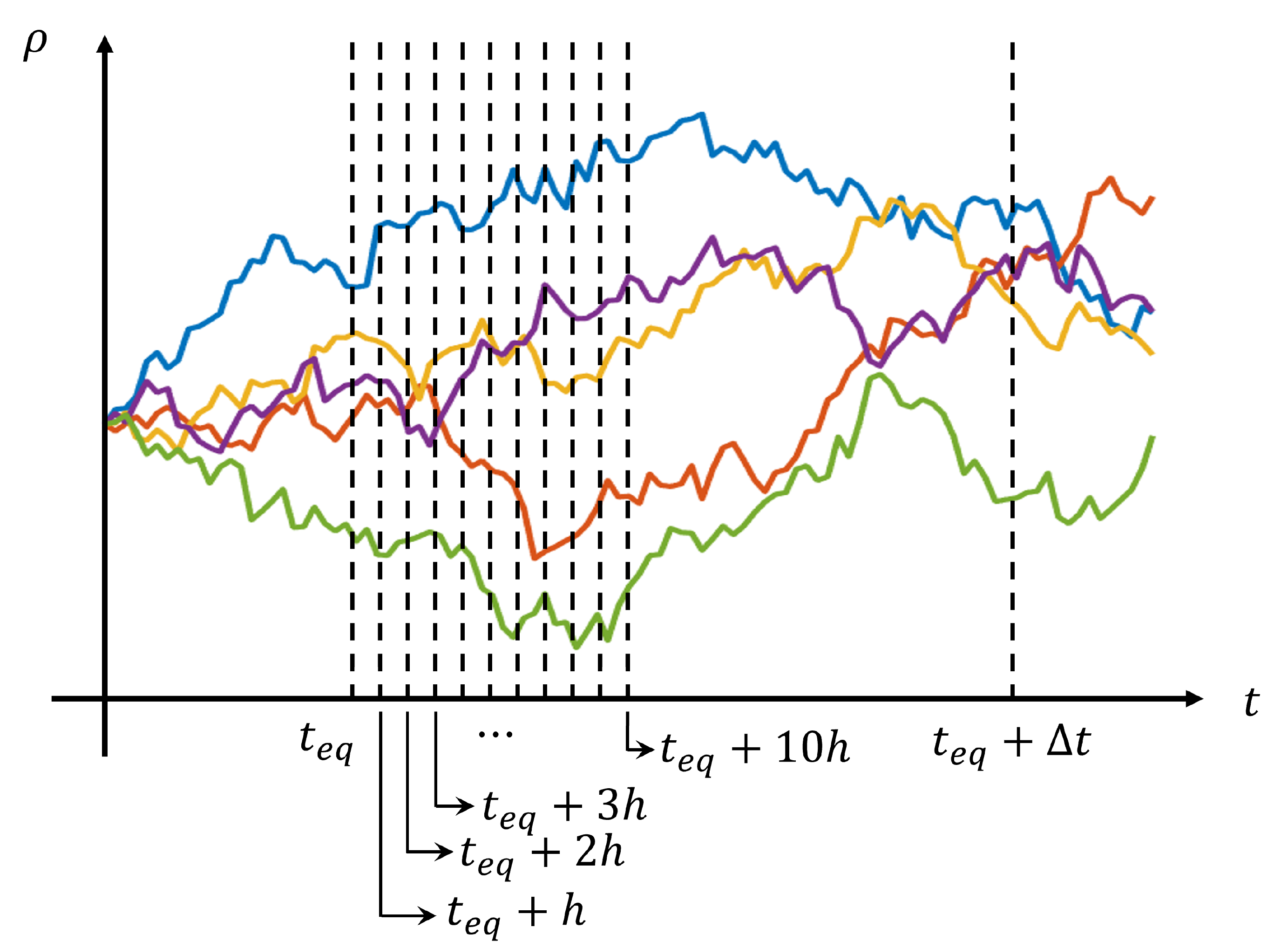}
\caption{Schematic for the data collection using Kinetic Monte-Carlo (KMC) simulations. For all cases in Sec.~\ref{Sec:Results}, we run $R=10^4$ realizations (five are shown in figure) for a given profile. In each realization, we first run a time interval $t_{eq}$ for achieving local equilibrium. The following 10 tiny timesteps $h$ are used for calculating the operator entries, and the density values at $t=t_{eq}$ and $t=t_{eq} + \Delta t$ are used for learning the free energy.}
\label{fig:KMC_Schematic}
\end{figure}

For the three cases shown in Sec.~\ref{Sec:Results}, all particle simulations used for data generation are run on a one dimensional system $x \in [0,0.5]$ with 2000 lattice sites, 25 shape functions and periodic boundary conditions, indicating a lattice size of $\epsilon = 2.5 \times 10^{-4}$ and a spatial discretization of $\Delta x_\gamma = 0.02$. We use 28 different cosine functions $\rho(x) = \rho_{ave} - A \cos (4 \pi f x)$ as initial profiles (see Table~\ref{Table:InitialProfile} for the choice of parameters $\rho_{ave}$, $A$ and $f$). The corresponding data distribution can found in Fig.~\ref{fig:DataDistribution}), which consists of $N_K = 28 \times 25 = 700$ data points. A schematic for the KMC simulation can be found in Fig.~\ref{fig:KMC_Schematic}. During the simulation, each profile is run for $R = 10^4$ realizations. Note that in each realization, since each lattice site can have at most one particle, the initial profile cannot match the targeted profile. The initial profile for each realization is generated according to the sample probability distribution such that the corresponding ensemble average over all realizations has the targeted cosine profile. Each realization is first run for time $t_{eq} = 50\epsilon^2\tau$ to achieve local equilibration, then following by $N_{\Delta t}$ number of time intervals $\Delta t$. Here we choose $N_{\Delta t} = 1$ for all three cases in Sec.~\ref{Sec:Results} and set $\Delta t$ as $20 \epsilon^2 \tau$, $40 \epsilon^2 \tau$ and $20 \epsilon^2 \tau$ for each case, respectively. The factor $\tau$ is used for compensating the timescale difference among different profiles. It is profile-related and is defined as $\tau=e^{2 J_0 L \rho_{\max}}$ with $\rho_{\max} = \rho_{ave} + A $ the maximum density in each profile. For each initial profile, the weak-form of the density for all $R$ realizations, $\langle \rho_\epsilon, \gamma_i \rangle$, are outputted at time $t=t_{eq} + n h$ with $n=0, 1, \cdots, N_h$. Here we choose $N_h = 10$, $h=0.01 \epsilon^2 \tau$. Then, profiles $\rho_{\epsilon,i}$ for all spatial indices $i$ are calculated by solving equations $\langle \rho_\epsilon, \gamma_i(x) \rangle = \sum_j \rho_{\epsilon,j} \langle \gamma_j, \gamma_i \rangle$. Given that $N_h h$ is extremely small from a macroscopic perspective such that the average profile barely changes during this interval, we regard these $R_K = N_h R = 10^5$ tiny time intervals $h$ as independent realizations \citep{embacher2018computing}. Next, Eq.~\ref{Eq:ComputeK} is applied to calculate the discrete operator entries $\left\langle \gamma_i, \K_z \gamma_i \right\rangle$ and $\left\langle \gamma_{i+1}, \K_z \gamma_i \right\rangle$ at all spatial points over $R_K$ realizations. This forms $N_K = 700$ number of training data points for both $K_0$ and $K_1$. 
Furthermore, the weak-form average profiles over $R$ realizations, $\langle \rho, \gamma_i \rangle$, at time $t=t_{eq} + n \Delta t$ with $n=0, 1, \cdots N_{\Delta t}$ are outputted, where $i=0, 1, \cdots, N_\gamma-1$, and the average profiles $\rho_i$ for all $i$ are obtained. This forms $N_f = N_K N_{\Delta t} = 700$ number of training data for learning the free energy according to the coarse-grained evolution equation during the short time interval $\Delta t$.

The particle simulations used for validation purposes are necessarily long in macroscopic time scales and thus extremely time consuming even after calculating each trajectory in parallel. In order to save computational cost, we choose to run the continuum evolution with initial profiles that include multiple periods under a periodic boundary condition but only run the simulation on a single period instead of the whole system. Here we use the triangular wave $\rho(x) = \rho_{ave} - A \left( \left|  
f x - \lfloor f x + 0.5 \rfloor  \right| - 1 \right) $ as initial profile, where $\rho_{ave}$, $A$ and $f$ represent the average density, amplitude and frequency, respectively, and $\lfloor a \rfloor$ is the floor function that outputs the greatest integer less or equal to $a$. The evolution within one period is duplicated to the full system ranging in $x\in[0,1]$. For example, for the two cases with $J_0 L= 0.9$ in Sec.~\ref{Sec:Results}, initial profiles with $f=2$ are used. Here particle simulations are run on $x\in[0,0.5]$ with 2000 lattice sites and 25 shape functions, which has the same spatial discretization at both the macroscopic scale $\Delta x_\gamma = 0.02$ and the microscopic scale $\epsilon=2.5\times 10^{-4}$ as the predictions from Stat-PINNs and PINNs.  For the two cases with $J_0 L= 2.2$, initial profiles with $f=3$ are used. The corresponding particle simulations are run on $x\in[0,1/3]$ with 1333 lattice sites and 17 shape functions, indicating $\Delta x_\gamma \approx 0.0196$ and $\epsilon \approx 2.5006\times 10^{-4}$, which is a good approximation of the spatial discretization of Stat-PINNs and PINNs. Notice that another rescaling trick is applied here for obtaining particle simulations within a spacial range different from $x \in [0,1]$ in order to reduce the complexity of the code. See Appendix~\ref{Sec:KMC_Rescaling} for more details.

\section{Rescaling for the Particle Simulations}
\label{Sec:KMC_Rescaling}

When bridging scales, a rescaling of length and time scales is usually required between the microscopic particle scale and the macroscopic continuum scale \citep{embacher2018computing}. 
In this section, we will introduce a rescaling trick for learning and predicting the dissipative evolution for systems with the same physics at the particle level but different spatial descriptions at the macroscale. 

First, we consider a target particle process on $x \in [0, \alpha]$ that has $N_b$ bins and $N_\gamma$ shape functions $\gamma_i(x)$. The Arrhenius process has an interaction strength of $J$ and an interaction range within $L_b$ neighbours on each side, indicating a length of interaction range of $L = \alpha L_b / N_b$. In the target process, the macroscopic times include $t_{eq}$, $h$ and $\Delta t$. The corresponding microscopic times are calculated by multiplying the macroscopic times by a factor $N_b^2 / \alpha^2$ \citep{embacher2018computing}. During postprocessing, the operator entries $K_{ij} = \langle \gamma_j, \K_\rho \gamma_i \rangle$ are calculated according to Eq.~\ref{Eq:ComputeK}, where $Y_{\gamma_i}$ is calculated as $Y_{\gamma_i} = \langle \rho - \bar{\rho}, \gamma_i \rangle / \sqrt{\Delta x_\gamma}$ with $\Delta x_\gamma = \alpha / N_\gamma$ and $\bar{\rho}$ as the average of $\rho$ over all realizations. Furthermore, using Eq.~\ref{Eq:Weak_T-GENERIC}, we can obtain the free energy $f(\rho)$ or $f(\rho, \nabla \rho)$.

Next, we consider a different system, which we will call the simulated system. This system also has $N_b$ bins and $N_\gamma$ shape functions but ranges between $x\in[0,1]$. We denote such shape functions as $\gamma_i^s(x)$ with a superscript $s$ to indicate that they refer to the  simulated system. Although $\gamma_i^s(x)$ is squeezed by a factor $\alpha$ with respect to $\gamma_i(x)$, it covers the same amount of bins as $\gamma_i(x)$ in the target system. Furthermore, the interaction strength $J$ and interaction range $L_b$ (in unit of bins) are the same as the ones in the target system, indicating the same Arrhenius process at the particle level. We note that the microscopic times for this simulated system should be calculated by multiplying the macroscopic times by a factor of $N_b^2$. Therefore, in order to keep the particle dynamics the same as that of the target system, all macroscopic times should be rescaled by multiplying by a factor $1/\alpha^2$ from the target system, i.e., $h^s = h / \alpha^2$ and $\Delta t^s = \Delta t / \alpha^2$. Given the same microscopic local dynamics, $\rho^s$ remains the same as $\rho$ since both are defined within the range $[0,1]$. Then, based on the rescaling of length and time scales, the following rescalings can be found, $\dot{\rho} = \dot{\rho}^s / \alpha^2$, $\nabla \rho = (\nabla \rho)^s / \alpha$, $M = \alpha M^s$ and $Y_{\gamma_i} = \sqrt{\alpha} Y^s_{\gamma^s_i}$. Therefore, the dissipative operator entries and free energy density for target system can be rescaled from the simulated system by $K_{ij} = K_{ij}^s / \alpha$, $f(\rho) = f^s(\rho)$ and $f(\rho, \nabla \rho) = f^s(\rho, \alpha \nabla \rho)$.

\begin{figure*}
\centering
\includegraphics[width=.85\linewidth]{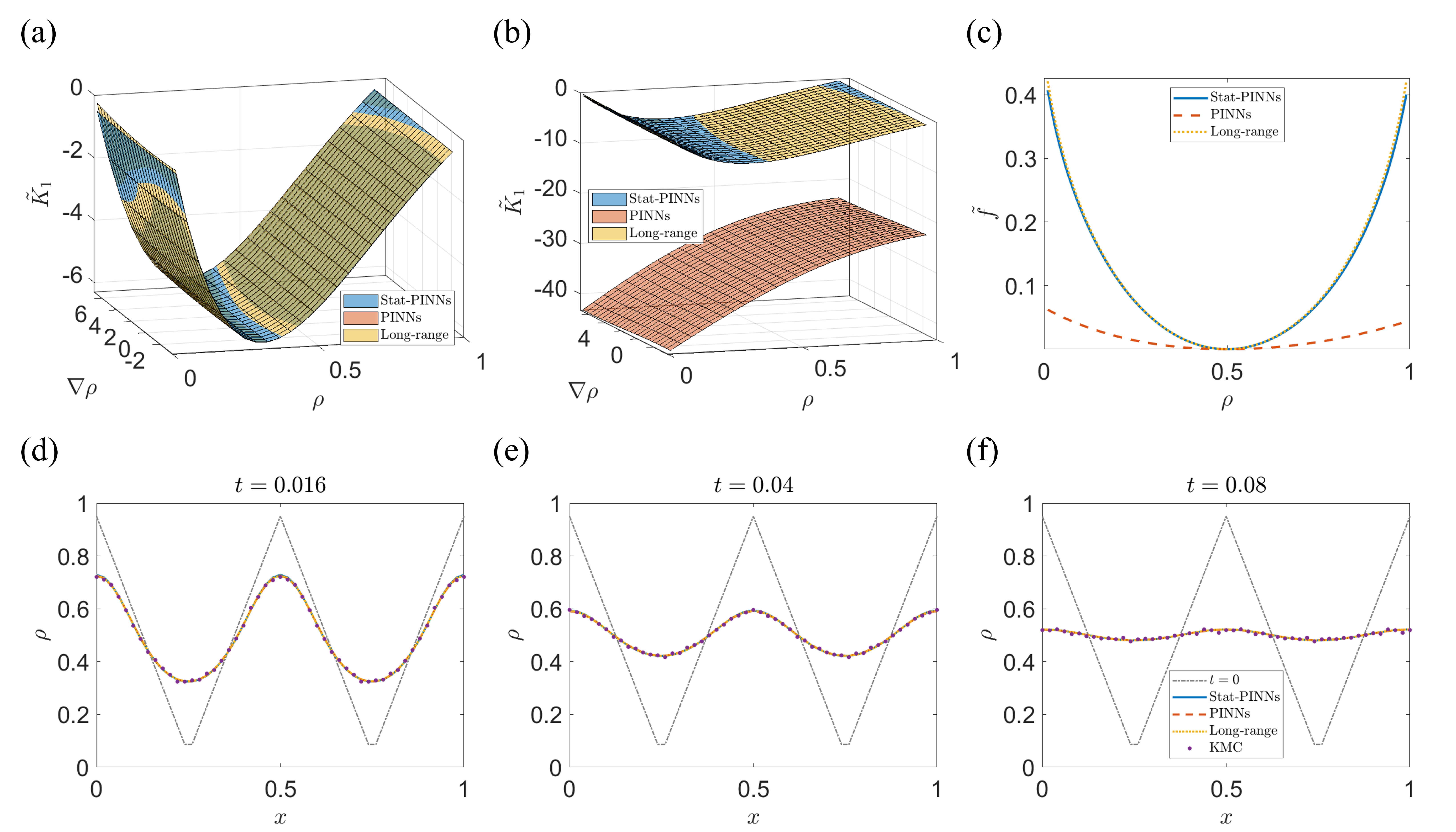}
\caption{Results for Arrhenius process with $JL=0.9, L=40\epsilon$ using additional training data, including (a, b) dissipative operator entry $\tilde{K}_1$ plotted in two different ranges, (c) calibrated free energy density $\tilde{f}$, and (d-f) snapshots of macroscopic evolution starting from a triangular wave initial profile (black dash dotted line). Predicting methods include Stat-PINNs (blue surfaces or blue solid lines), PINNs (orange surfaces or orange dashed lines) and long-range analytic model (yellow surfaces or yellow dotted lines). Results from KMC particle simulations (purple dots) are used as true macroscopic evolution for comparison.}
\label{fig:JL0.9_L40_LargeData}
\end{figure*}

\section{Settings for Stat-PINNs and PINNs Training}
\label{Append:Training_Settings}

Stat-PINNs involve two sequential training processes for $\text{NN}_{K_1}$ and $\text{NN}_f$. All neural networks are fully-connected feed-forward networks with SoftPlus activation functions and are implemented in Python using JAX as the main library. Adams optimizer with a learning rate of $10^{-2}$ is used for all cases in Stat-PINNs. For all cases in Sec.~\ref{Sec:Results}, $\text{NN}_{K_1}$ has 2 hidden layers with 20 neurons per layer and is trained for 6000 epochs. $\text{NN}_f$ has 2 hidden layers with 20 neurons per layer and is trained for 2000 epochs. 

All cases in Sec~\ref{Sec:Results} are also trained by PINNs as comparison, which also consist of $\text{NN}_{K_1}$ and $\text{NN}_f$ but are trained jointly using the following loss function 
\begin{equation}
\begin{split}
\label{Eq:Loss_PINNs}
    \mathcal{L}_{\text{PINNs}}
    & = \frac{1}{2N_f} \sum_{s=1}^{N_f} 
    \left\| \sum_{i} \langle \gamma_{j^s}, \gamma_{i} \rangle
    \frac{\Delta z_i^{(s)}}{\Delta t}
    \right.
    \\
    & \quad 
    \left.
    + \sum_{i} \langle \gamma_{j^s}, \K_{z^{(s)}} \gamma_i \rangle 
    Q(\mathbf{Z}^{Q(s)}_{i}; \boldsymbol{\theta}_f)
    \right\|^2.
\end{split}
\end{equation}
This is almost the same loss function as the one used for $\text{NN}_f$ in Stat-PINNs (Eq.~\ref{Eq:Loss_feNN}) but setting the variance on denominator to one. Notice that now the operator entries $\langle \gamma_{j^s}, \K_{z^{(s)}} \gamma_i \rangle$ are represented by $\text{NN}_{K_1}$ to be trained. Structures of $\text{NN}_{K_1}$ and $\text{NN}_f$ in PINNs are identical as those in Stat-PINNs, but parameters $\boldsymbol{\theta}_{1}$ and $\boldsymbol{\theta}_f$ are trained simultaneously. All cases are trained for 10000 epochs using Adams optimizer with learning rate of $10^{-3}$. We remark that for both Stat-PINNs and PINNs, the loss functions have reached stable plateaus after the epochs previously mentioned, indicating the convergence of the training process, though the results from PINNs may not be robust enough. 

\section{Non-Uniqueness Problem for the Free Energy}
\label{Append:fe_non-unique}

As discussed in Sec.~\ref{Sec:fe_Discrete}, different functional dependencies of the free energy density $f$ lead to different forms of the driving force and hence the evolution equations. Due to the mathematical structure of the evolution equations, $f$ cannot be uniquely determined from the macroscopic evolution equation, even when the operator is uniquely identified. In this section, we will discuss the non-uniqueness of the free energy based for the three different cases. For simplicity, we will use $\rho$ as the field variable in this section to keep it the same as in the Arrhenius example.

\begin{figure*}
\centering
\includegraphics[width=.85\linewidth]{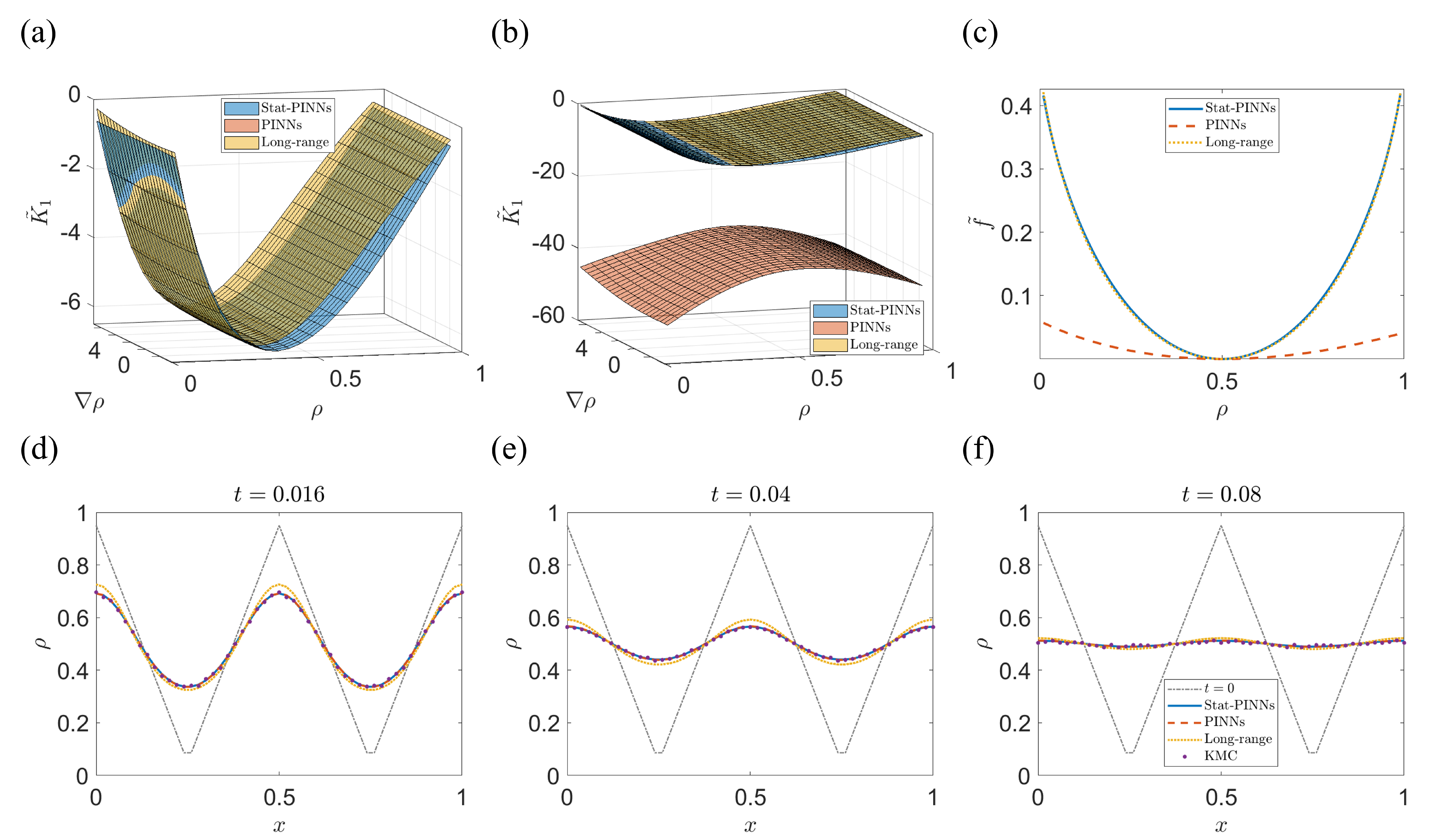}
\caption{Results for Arrhenius process with $JL=0.9, L=2\epsilon$ using additional training data, including (a, b) dissipative operator entry $\tilde{K}_1$ plotted in two different ranges, (c) calibrated free energy density $\tilde{f}$, and (d-f) snapshots of macroscopic evolution starting from a triangular wave initial profile (black dash dotted line). Predicting methods include Stat-PINNs (blue surfaces or blue solid lines), PINNs (orange surfaces or orange dashed lines) and long-range analytic model (yellow surfaces or yellow dotted lines). Results from KMC particle simulations (purple dots) are used as true macroscopic evolution for comparison.}
\label{fig:JL0.9_L2_LargeData}
\end{figure*}

First, we assume that the free energy density can be written as $f=f(\rho)$. This is the simplest case and also the one used in Sec.~\ref{Sec:Results}. In this case, the thermodynamic force is given by Eq.~\ref{Eq:Q_f(z)} as $Q=f'(\rho)$ and the governing equation Eq.~\ref{Eq:Weak_T-GENERIC_MassCons} is a diffusion equation. The spatial derivative involved in the operator $\K_\rho$ indicates that adding a constant to $Q$ does not lead to any difference in the evolution. Therefore, $f(\rho)$ is unique up to an arbitrary linear function. Here we introduce the calibrated free energy density defined as
\begin{equation}
\label{Eq:cali_f(rho)}
    \tilde{f}(\rho) = f(\rho) - f(\rho_{r0}) - f'(\rho_{r1}) \left( \rho - \rho_{r0} \right).
\end{equation}
which can be understood as forcing $\tilde{f}(\rho_{r0}) = 0$ and $\tilde{f}'(\rho_{r1})=0$ at two reference points $\rho_{r0}$ and $\rho_{r1}$. For simplicity, we choose $\rho_{r0} = \rho_{r1}=0.5$ as reference points. We acknowledge that defining $f(0)=0$ would be more natural from a physical standpoint as there is no energy when no particle exists. $\rho_{r0}=0.5$ is chosen for practical reasons, to facilitate  plotting and comparing free energy profiles from different models in a easier and more intuitive way.

Second, when assuming $f=f(\rho, \nabla \rho)$, the thermodynamic force is given by Eq.~\ref{Eq:Q_f(z,grad_z)} as $Q = f_{,\rho} - \nabla \cdot f_{,\nabla \rho}$ and the governing equation is a fourth order equation, such as the Cahn–Hilliard equation. Similarly to the case above, $Q$ is unique up to an arbitrary constant. 
We define the calibrated free energy density as
\begin{equation}
\begin{split}
\label{Eq:cali_f(rho,grad)}
    \tilde{f}(\rho, \nabla \rho) 
    & = f(\rho, \nabla \rho)
    - f(\rho_{r0}, 0) 
    \\
    & \quad 
    - f_{,\rho} (\rho_{r1}, 0)
    \left( \rho - \rho_{r0} \right)
    - f_{, \nabla \rho} (\rho, 0) \nabla \rho,
\end{split}
\end{equation}
which satisfies $f(\rho_{r0},0)=0$, $f_{,\rho}(\rho_{r1},0) = 0$ and $f_{,\nabla \rho}(\rho,0) = 0$. Similar to the case above, we choose $\rho_{r0} = \rho_{r1} = 0.5$  as reference points.

The final case to discuss is $f=f(\rho, \nabla^2 \rho)$, which corresponds to the long-range analytic model for Arrhenius process when the convolution is approximated to second order as $J\ast \rho \approx 2 J_0 L \left( \rho + L^2 \nabla^2 \rho / 6 \right)$. In this case, the thermodynamic force is given by $Q = f_{,\rho} + \nabla^2 f_{,\nabla^2 \rho}$ according to Eq.~\ref{Eq:Q_f(z,grad_z,lap_z)}. 
We define the calibrated free energy density as
\begin{equation}
\begin{split}
\label{Eq:cali_f(rho,lap)}
    \tilde{f}(\rho, \nabla^2 \rho) 
    & = f(\rho, \nabla^2 \rho)
    - f(\rho_{r0}, 0) 
    \\
    & 
    - f_{,\rho} (\rho_{r1}, 0)
    \left( \rho - \rho_{r0} \right)
    - f_{, \nabla^2 \rho} (0, 0) \nabla^2 \rho,
\end{split}
\end{equation}
which satisfies $f(\rho_{r0},0)=0$, $f_{,\rho}(\rho_{r1},0) = 0$ and $f_{,\nabla^2 \rho}(0,0) = 0$. Here we choose $\rho_{r0} = \rho_{r1} = 0.5$ as reference points.

\section{Results with Sufficient Training Data}
\label{Append:LargeData}

\begin{figure*}
\centering
\includegraphics[width=.85\linewidth]{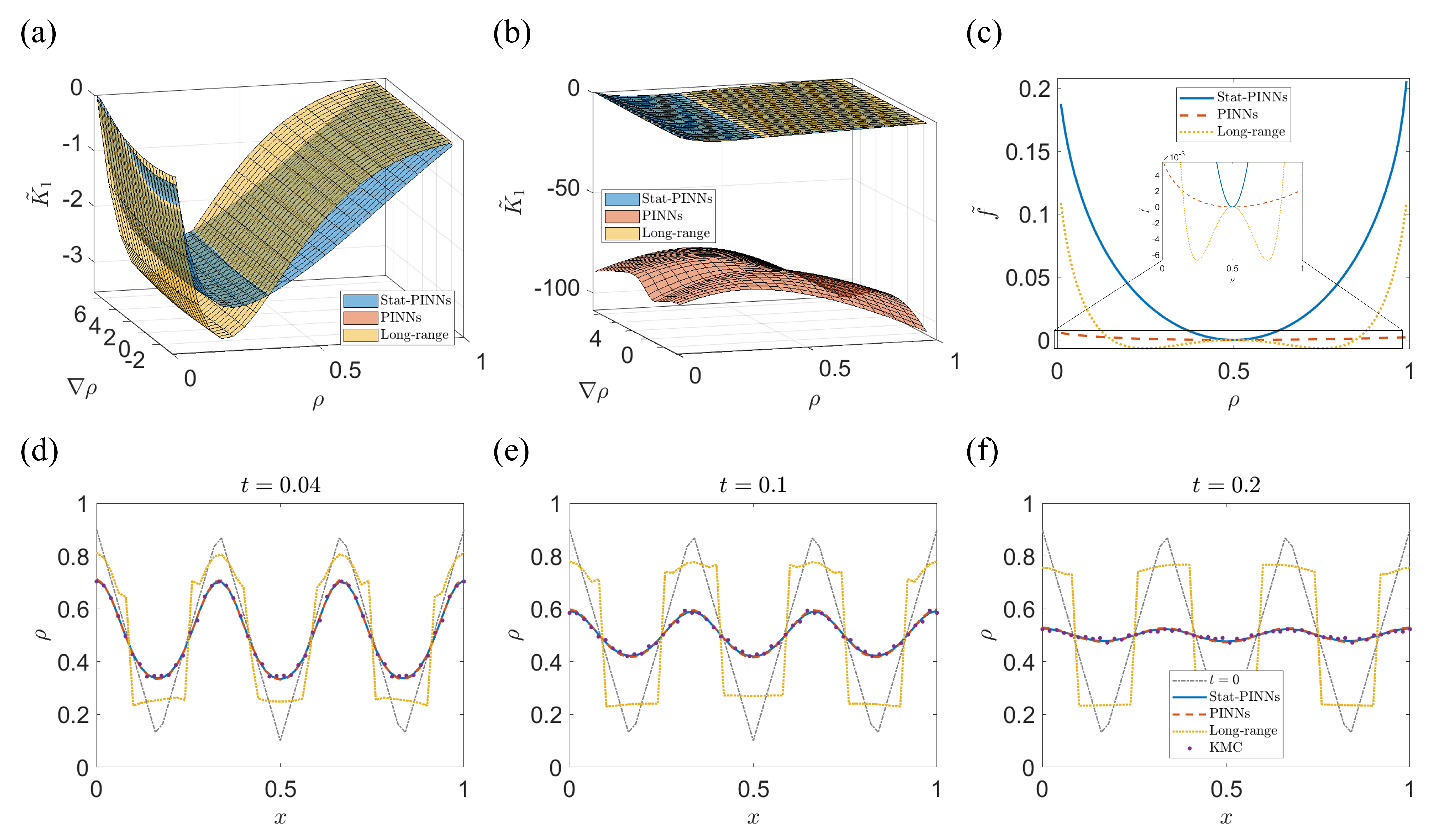}
\caption{Results for Arrhenius process with $JL=2.2, L=2\epsilon$ using sufficient training data, including (a, b) dissipative operator entry $\tilde{K}_1$ plotted in two different ranges, (c) calibrated free energy density $\tilde{f}$, and (d-f) snapshots of macroscopic evolution starting from a triangular wave initial profile (black dash dotted line). Predicting methods include Stat-PINNs (blue surfaces or blue solid lines), PINNs (orange surfaces or orange dashed lines) and long-range analytic model (yellow surfaces or yellow dotted lines). Results from KMC particle simulation (purple dots) are used as true macroscopic evolution for comparison.}
\label{fig:JL2.2_L2_LargeData}
\end{figure*}

While Stat-PINNs successfully predict the continuum evolution for the first three cases in Sec.~\ref{Sec:Results}, results from PINNs are not as good with the given limited data. In this section, we use almost the same parameter setting as in Appendix~\ref{Append:KMC_Settings}, but increase the training data for $\text{NN}_f$ by 10 times, i.e.,  $N_{\Delta t}=10$ and $N_f = 7000$. Results of the off-diagonal operator entry $K_1$, the calibrated free energy density $\tilde{f}$ and the continuum evolution for all three cases in Sec.~\ref{Sec:Results} are shown in Figs.~\ref{fig:JL0.9_L40_LargeData}--\ref{fig:JL2.2_L2_LargeData}. As we can see, although PINNs cannot predict the true operator or free energy, the predicted continuum evolution from PINNs now match the true results from particle simulations. This indicates that PINNs can predict the continuum evolution as long as there exists sufficient training data such that the noise from stochastic particle dynamics is negligible, as could be expected. However, as noted continuum data alone, in the absence of statistical physics, is insufficient to uniquely characterize the free energy and operator. 




\bibliographystyle{plainnat}
\bibliography{Reference}

\begin{thebibliography}{38}
\providecommand{\natexlab}[1]{#1}
\providecommand{\url}[1]{\texttt{#1}}
\expandafter\ifx\csname urlstyle\endcsname\relax
  \providecommand{\doi}[1]{doi: #1}\else
  \providecommand{\doi}{doi: \begingroup \urlstyle{rm}\Url}\fi

\bibitem[Adams et~al.(2013)Adams, Dirr, Peletier, and Zimmer]{adams2013large}
Stefan Adams, Nicolas Dirr, Mark Peletier, and Johannes Zimmer.
\newblock Large deviations and gradient flows.
\newblock \emph{Philosophical Transactions of the Royal Society A:
  Mathematical, Physical and Engineering Sciences}, 371\penalty0
  (2005):\penalty0 20120341, 2013.

\bibitem[Arampatzis et~al.(2011)Arampatzis, Katsoulakis, Plechá{\v{c}},
  Taufer, and Xu]{Arampatzis2011Hierarchical}
Giorgos Arampatzis, Markos~A. Katsoulakis, Petr Plechá{\v{c}}, Michela Taufer,
  and Lifan Xu.
\newblock Hierarchical fractional-step approximations and parallel kinetic
  {M}onte {C}arlo algorithms.
\newblock \emph{Journal of Computational Physics}, 231\penalty0 (23):\penalty0
  7795--7814, 2011.

\bibitem[Bodineau et~al.(2016)Bodineau, Gallagher, and
  Saint-Raymond]{bodineau2016brownian}
Thierry Bodineau, Isabelle Gallagher, and Laure Saint-Raymond.
\newblock The {B}rownian motion as the limit of a deterministic system of
  hard-spheres.
\newblock \emph{Inventiones Mathematicae}, 203:\penalty0 493--553, 2016.

\bibitem[Bortz and Lebowitz(1975)]{bortz1975new}
Malvin~H. Bortz, Alfred B.~and~Kalos and Joel~L. Lebowitz.
\newblock A new algorithm for {M}onte {C}arlo simulation of {I}sing spin
  systems.
\newblock \emph{Journal of Computational Physics}, 17\penalty0 (1):\penalty0
  10--18, 1975.

\bibitem[Dewey(1994)]{dewey1994arrhenius}
William~C. Dewey.
\newblock Arrhenius relationships from the molecule and cell to the clinic.
\newblock \emph{International Journal of Hyperthermia}, 10\penalty0
  (4):\penalty0 457--483, 1994.

\bibitem[Dietrich et~al.(2023)Dietrich, Makeev, Kevrekidis, Evangelou,
  Bertalan, Reich, and Kevrekidis]{dietrich2023learning}
Felix Dietrich, Alexei Makeev, George Kevrekidis, Nikolaos Evangelou, Tom
  Bertalan, Sebastian Reich, and Ioannis~G. Kevrekidis.
\newblock Learning effective stochastic differential equations from microscopic
  simulations: Linking stochastic numerics to deep learning.
\newblock \emph{Chaos: An Interdisciplinary Journal of Nonlinear Science},
  33\penalty0 (2), 2023.

\bibitem[Embacher et~al.(2018)Embacher, Dirr, Zimmer, and
  Reina]{embacher2018computing}
Peter Embacher, Nicolas Dirr, Johannes Zimmer, and Celia Reina.
\newblock Computing diffusivities from particle models out of equilibrium.
\newblock \emph{Proceedings of the Royal Society A: Mathematical, Physical and
  Engineering Sciences}, 474\penalty0 (2212):\penalty0 20170694, 2018.

\bibitem[Fehrman and Gess(2021)]{Fehrman2021a}
Benjamin Fehrman and Benjamin Gess.
\newblock Well-posedness of the {D}ean-{K}awasaki and the nonlinear
  {D}awson-{W}atanabe equation with correlated noise, 2021.

\bibitem[Gibbs(1972)]{gibbs1972sufficient}
Julian~H. Gibbs.
\newblock Sufficient conditions for the {A}rrhenius rate law.
\newblock \emph{The Journal of Chemical Physics}, 57\penalty0 (10):\penalty0
  4473--4478, 1972.

\bibitem[Gilmer and Bennema(1972)]{gilmer1972simulation}
G.H. Gilmer and P.~Bennema.
\newblock Simulation of crystal growth with surface diffusion.
\newblock \emph{Journal of Applied Physics}, 43\penalty0 (4):\penalty0
  1347--1360, 1972.

\bibitem[Grmela and {\"O}ttinger(1997)]{grmela1997dynamics}
Miroslav Grmela and Hans~Christian {\"O}ttinger.
\newblock Dynamics and thermodynamics of complex fluids. {I}. {D}evelopment of
  a general formalism.
\newblock \emph{Physical Review E}, 56\penalty0 (6):\penalty0 6620, 1997.

\bibitem[Hernandez et~al.(2021)Hernandez, Badias, Gonzalez, Chinesta, and
  Cueto]{hernandez2021deep}
Quercus Hernandez, Alberto Badias, David Gonzalez, Francisco Chinesta, and
  Elias Cueto.
\newblock Deep learning of thermodynamics-aware reduced-order models from data.
\newblock \emph{Computer Methods in Applied Mechanics and Engineering},
  379:\penalty0 113763, 2021.

\bibitem[Huang et~al.(2020)Huang, Aguado-Montero, Dirr, Zimmer, and
  Reina]{huang2020particle}
Shenglin Huang, Santiago Aguado-Montero, Nicolas Dirr, Johannes Zimmer, and
  Celia Reina.
\newblock From particle fluctuations to macroscopic evolution equations: The
  case of exclusion dynamics.
\newblock \emph{CEUR Workshop Proceedings}, 2783:\penalty0 140--152, 2020.

\bibitem[Huang et~al.(2022)Huang, He, and Reina]{huang2022variational}
Shenglin Huang, Zequn He, and Celia Reina.
\newblock Variational {O}nsager {N}eural {N}etworks ({VONN}s): A
  thermodynamics-based variational learning strategy for non-equilibrium pdes.
\newblock \emph{Journal of the Mechanics and Physics of Solids}, 163:\penalty0
  104856, 2022.

\bibitem[Katsoulakis and Vlachos(2003)]{katsoulakis2003coarse}
Markos~A. Katsoulakis and Dionisios~G. Vlachos.
\newblock Coarse-grained stochastic processes and kinetic {M}onte {C}arlo
  simulators for the diffusion of interacting particles.
\newblock \emph{The Journal of Chemical Physics}, 119\penalty0 (18):\penalty0
  9412--9427, 2003.

\bibitem[Kipnis and Landim(1998)]{kipnis1998scaling}
Claude Kipnis and Claudio Landim.
\newblock \emph{Scaling limits of interacting particle systems}, volume 320.
\newblock Springer Science \& Business Media, 1998.

\bibitem[Konarovskyi et~al.(2019)Konarovskyi, Lehmann, and von
  Renesse]{Renesse}
Vitalii Konarovskyi, Tobias Lehmann, and Max-K. von Renesse.
\newblock Dean-{K}awasaki dynamics: ill-posedness vs. triviality.
\newblock \emph{Electron. Commun. Probab.}, 24:\penalty0 Paper No. 8, 9, 2019.
\newblock ISSN 1083-589X.
\newblock \doi{10.1214/19-ECP208}.
\newblock URL \url{https://doi.org/10.1214/19-ECP208}.

\bibitem[Kraaij et~al.(2020)Kraaij, Lazarescu, Maes, and
  Peletier]{kraaij2020fluctuation}
Richard~C. Kraaij, Alexandre Lazarescu, Christian Maes, and Mark Peletier.
\newblock Fluctuation symmetry leads to {GENERIC} equations with non-quadratic
  dissipation.
\newblock \emph{Stochastic Processes and their Applications}, 130\penalty0
  (1):\penalty0 139--170, 2020.

\bibitem[Kubo(1966)]{kubo1966fluctuation}
Rep Kubo.
\newblock The fluctuation-dissipation theorem.
\newblock \emph{Reports on Progress in Physics}, 29\penalty0 (1):\penalty0 255,
  1966.

\bibitem[Laidler(1984)]{laidler1984development}
Keith~J. Laidler.
\newblock The development of the {A}rrhenius equation.
\newblock \emph{Journal of Chemical Education}, 61\penalty0 (6):\penalty0 494,
  1984.

\bibitem[Lee et~al.(2021)Lee, Trask, and Stinis]{lee2021machine}
Kookjin Lee, Nathaniel Trask, and Panos Stinis.
\newblock Machine learning structure preserving brackets for forecasting
  irreversible processes.
\newblock \emph{Advances in Neural Information Processing Systems},
  34:\penalty0 5696--5707, 2021.

\bibitem[Leli{\`e}vre et~al.(2010)Leli{\`e}vre, Rousset, and
  Stoltz]{Lelievre2010a}
Tony Leli{\`e}vre, Mathias Rousset, and Gabriel Stoltz.
\newblock \emph{Free energy computations}.
\newblock Imperial College Press, London, 2010.
\newblock ISBN 978-1-84816-247-1; 1-84816-247-2.
\newblock \doi{10.1142/9781848162488}.
\newblock URL \url{http://dx.doi.org/10.1142/9781848162488}.
\newblock A mathematical perspective.

\bibitem[Li et~al.(2019)Li, Dirr, Embacher, Zimmer, and
  Reina]{li2019harnessing}
Xiaoguai Li, Nicolas Dirr, Peter Embacher, Johannes Zimmer, and Celia Reina.
\newblock Harnessing fluctuations to discover dissipative evolution equations.
\newblock \emph{Journal of the Mechanics and Physics of Solids}, 131:\penalty0
  240--251, 2019.

\bibitem[Linderoth et~al.(1997)Linderoth, Horch, L{\ae}gsgaard, Stensgaard, and
  Besenbacher]{linderoth1997surface}
Trolle~Ren{\'e} Linderoth, Sebastian Horch, Erik L{\ae}gsgaard, Ivan
  Stensgaard, and Flemming Besenbacher.
\newblock Surface diffusion of {P}t on {P}t (110): Arrhenius behavior of long
  jumps.
\newblock \emph{Physical Review Letters}, 78\penalty0 (26):\penalty0 4978,
  1997.

\bibitem[Mielke(2011)]{mielke2011formulation}
Alexander Mielke.
\newblock Formulation of thermoelastic dissipative material behavior using
  {GENERIC}.
\newblock \emph{Continuum Mechanics and Thermodynamics}, 23\penalty0
  (3):\penalty0 233--256, 2011.

\bibitem[Montefusco et~al.(2021)Montefusco, Peletier, and
  {\"O}ttinger]{montefusco2021framework}
Alberto Montefusco, Mark~A Peletier, and Hans~Christian {\"O}ttinger.
\newblock A framework of nonequilibrium statistical mechanics. {II}.
  {C}oarse-graining.
\newblock \emph{Journal of Non-Equilibrium Thermodynamics}, 46\penalty0
  (1):\penalty0 15--33, 2021.

\bibitem[{\"O}ttinger(2005)]{ottinger2005beyond}
Hans~Christian {\"O}ttinger.
\newblock \emph{Beyond equilibrium thermodynamics}.
\newblock John Wiley \& Sons, 2005.

\bibitem[{\"O}ttinger and Grmela(1997)]{ottinger1997dynamics}
Hans~Christian {\"O}ttinger and Miroslav Grmela.
\newblock Dynamics and thermodynamics of complex fluids. {II}. {I}llustrations
  of a general formalism.
\newblock \emph{Physical Review E}, 56\penalty0 (6):\penalty0 6633, 1997.

\bibitem[{\"O}ttinger et~al.(2021){\"O}ttinger, Peletier, and
  Montefusco]{ottinger2021framework}
Hans~Christian {\"O}ttinger, Mark~A. Peletier, and Alberto Montefusco.
\newblock A framework of nonequilibrium statistical mechanics. {I}. {R}ole and
  types of fluctuations.
\newblock \emph{Journal of Non-Equilibrium Thermodynamics}, 46\penalty0
  (1):\penalty0 1--13, 2021.

\bibitem[Presutti(2009)]{Presutti2009a}
Errico Presutti.
\newblock \emph{Scaling limits in statistical mechanics and microstructures in
  continuum mechanics}.
\newblock Theoretical and Mathematical Physics. Springer, Berlin, 2009.
\newblock ISBN 978-3-540-73304-1.

\bibitem[Schulze(2008)]{Schulze2008Efficient}
Tim~P. Schulze.
\newblock Efficient kinetic {M}onte {C}arlo simulation.
\newblock \emph{Journal of Computational Physics}, 227\penalty0 (4):\penalty0
  2455--2462, 2008.

\bibitem[Seifert(2012)]{seifert2012stochastic}
Udo Seifert.
\newblock Stochastic thermodynamics, fluctuation theorems and molecular
  machines.
\newblock \emph{Reports on progress in physics}, 75\penalty0 (12):\penalty0
  126001, 2012.

\bibitem[Sevick et~al.(2008)Sevick, Prabhakar, Williams, and
  Searles]{sevick2008fluctuation}
Edith~M. Sevick, Ranganathan Prabhakar, Stephen~R. Williams, and Debra~J.
  Searles.
\newblock Fluctuation theorems.
\newblock \emph{Annu. Rev. Phys. Chem.}, 59:\penalty0 603--633, 2008.

\bibitem[Sivaprasad et~al.(2021)Sivaprasad, Singh, Manwani, and
  Gandhi]{sivaprasad2021curious}
Sarath Sivaprasad, Ankur Singh, Naresh Manwani, and Vineet Gandhi.
\newblock The curious case of convex neural networks.
\newblock In \emph{Machine Learning and Knowledge Discovery in Databases.
  Research Track: European Conference, ECML PKDD 2021, Bilbao, Spain, September
  13--17, 2021, Proceedings, Part I 21}, pages 738--754. Springer, 2021.

\bibitem[Vlachos and Katsoulakis(2000)]{vlachos2000derivation}
D.G. Vlachos and M.A. Katsoulakis.
\newblock Derivation and validation of mesoscopic theories for diffusion of
  interacting molecules.
\newblock \emph{Physical Review Letters}, 85\penalty0 (18):\penalty0 3898,
  2000.

\bibitem[Wang et~al.(2022)Wang, Yu, and Perdikaris]{wang2022and}
Sifan Wang, Xinling Yu, and Paris Perdikaris.
\newblock When and why pinns fail to train: A neural tangent kernel
  perspective.
\newblock \emph{Journal of Computational Physics}, 449:\penalty0 110768, 2022.

\bibitem[Zhang et~al.(2022)Zhang, Shin, and Em~Karniadakis]{zhang2022gfinns}
Zhen Zhang, Yeonjong Shin, and George Em~Karniadakis.
\newblock {GFINN}s: {GENERIC} formalism informed neural networks for
  deterministic and stochastic dynamical systems.
\newblock \emph{Philosophical Transactions of the Royal Society A},
  380\penalty0 (2229):\penalty0 20210207, 2022.

\bibitem[Zhdanov(1991)]{zhdanov1991arrhenius}
Vladimir~Petrovich Zhdanov.
\newblock Arrhenius parameters for rate processes on solid surfaces.
\newblock \emph{Surface Science Reports}, 12\penalty0 (5):\penalty0 185--242,
  1991.

\end{thebibliography}

\end{document}